\documentclass[usenatbib]{mn2e}
\usepackage{amsmath, amssymb}
\usepackage{amsfonts}
\usepackage{epsfig,rotating}
\usepackage[T1]{fontenc}
\usepackage{aecompl}

\def\simeq{
\mathrel{\raise.3ex\hbox{$\sim$}\mkern-14mu\lower0.4ex\hbox{$-$}}
}

\def\ltsima{$\; \buildrel < \over \sim \;$}
\def\simlt{\lower.5ex\hbox{\ltsima}}
\def\gtsima{$\; \buildrel > \over \sim \;$}
\def\simgt{\lower.5ex\hbox{\gtsima}}

\def\msun{{\rm M_{\odot}}}

\def\be{\begin{equation}}
\def\ee{\end{equation}}

\def\del#1{{}}
\def\ltsima{$\; \buildrel < \over \sim \;$}
\def\simlt{\lower.5ex\hbox{\ltsima}}
\def\gtsima{$\; \buildrel > \over \sim \;$}
\def\simgt{\lower.5ex\hbox{\gtsima}}

\newcommand{\apj}{ApJ}

\newcommand{\mnras}{MNRAS}
\newcommand{\aap}{A\&A}

\newcommand{\apjl}{ApJL}

\newcommand{\nat}{Nature}

\title[A simple way to improve AGN feedback prescription]{A simple way to improve AGN feedback prescription in SPH simulations}
\author[Kastytis Zubovas, Martin A. Bourne, Sergei Nayakshin]{Kastytis Zubovas$^{1,\star}$, Martin A. Bourne$^{2,3}$ and Sergei Nayakshin$^{2}$ \\
  $^{1}$Center for Physical Sciences and Technology, Savanori\c{u} 231, Vilnius LT-02300, Lithuania \\
  $^{2}$Department of Physics \& Astronomy, University of Leicester, Leicester, LE1 7RH, UK \\
  $^{3}$Institute of Astronomy and Kavli Institute for Cosmology, University of Cambridge, Madingley Road, Cambridge, CB3 0HA, UK \\
  $^{\star}$ {E-mail:~} {\rm kastytis.zubovas@ftmc.lt} }

\begin{document}

\maketitle

\begin{abstract}
AGN feedback is an important ingredient in galaxy evolution, however
its treatment in numerical simulations is necessarily approximate,
requiring subgrid prescriptions due to the dynamical range involved in
the calculations. We present a suite of SPH simulations designed to
showcase the importance of the choice of a particular subgrid
prescription for AGN feedback. We concentrate on two approaches to
treating wide-angle AGN outflows: thermal feedback, where thermal and
kinetic energy is injected into the gas surrounding the SMBH particle,
and virtual particle feedback, where energy is carried by tracer
particles radially away from the AGN. We show that the latter model
produces a far more complex structure around the SMBH, which we argue
is a more physically correct outcome. We suggest a simple improvement
to the thermal feedback model - injecting the energy into a cone,
rather than spherically symmetrically - and show that this markedly
improves the agreement between the two prescriptions, without
requiring any noticeable increase in the computational cost of the
simulation.
\end{abstract}


\section{Introduction}

Feedback from Active Galactic Nuclei (AGN) is a key ingredient in
modern galaxy evolution models. It is required in order to explain the
sharp drop-off in the galaxy mass function above $M_* \simeq 10^{11}
\msun$, prevent the cooling catastrophe in galaxy clusters and produce
the hot gas atmospheres seen around many galaxies. Observations of
massive kpc-scale outflows \citep{Feruglio2010A&A, Sturm2011ApJ,
  Rupke2011ApJ, Cicone2014A&A} and pc-scale relativistic winds
\citep{Tombesi2010A&A, Tombesi2010ApJ} provide further evidence that
AGN affect their host galaxies in a significant way.

One outstanding issue in understanding the precise effects of feedback
is the range of spatial scales over which it operates. The AGN jets
and winds are launched from the accretion disc on scales $l_{\rm min}
< 0.01$~pc, while the observed effects on host galaxies span $10^3$~pc
or more. This means that in order to model the AGN feedback precisely,
a simulation should span at least 5 orders of magnitude in linear
scale, resulting in $10^{15}$ resolution elements for a 3D model, a
resolution which is not likely to be reached any time soon. Even
resolving the interaction of an AGN wind with the host galaxy
material, on scales of several parsecs and larger, is currently only
possible in single-galaxy models, rather than cosmological
simulations \citep{Schaye2015MNRAS}.

As a result of this shortcoming, some effects of AGN feedback might
not be apparent from simulations. In a recent paper
\citep{Bourne2015MNRAS}, we showed that mass resolution has a
significant influence upon the AGN feedback effects seen in
simulations of turbulent gas. In particular, at mass resolutions
typical of cosmological simulations, we found AGN feedback to be much
more negative, in the sense of efficiency of gas removal, than at
higher resolution. In high-resolution simulations, AGN winds expel
only diffuse gas, while denser gas forms filaments which fall in
toward the central supermassive black hole (SMBH) and can potentially
feed it or form stars. These results suggest that AGN feedback can be
positive as well as negative, increasing the star formation rate in
its host. Our other previous research \citep{Nayakshin2012MNRASb,
  Zubovas2013MNRAS, Bourne2014MNRAS} provides some support for this
conclusion as well. Several other authors have investigated similar
positive AGN feedback coming from the interaction of the jet with the
galactic ISM \citep{Silk2005MNRAS, Gaibler2012MNRAS}.

In this paper, we investigate another numerical aspect of modelling
AGN feedback: the prescription detailing how the AGN feedback energy
is passed to the surrounding gas. We perform high-resolution
simulations of AGN feedback acting upon a turbulent gas reservoir,
passing the feedback energy in one of two ways: a Monte Carlo
radiative transfer scheme, which self-consistently takes into account
gas optical depth; and a kernel-weighted thermal and kinetic energy
input scheme. We find that the Monte Carlo scheme, while being
numerically far more expensive and difficult to scale, provides
qualitatively different results than thermal and kinetic energy
injection. In particular, the Monte Carlo scheme allows the formation
of simultaneous gas inflows and outflows, resulting in prolonged AGN
feeding and faster, but less massive, outflows. We also test an
adjusted version of both schemes, where the feedback energy is
injected, or virtual particles released, biconically, rather than
spherically symmetrically. This change makes the results of the two
schemes significantly more alike. We suggest that using this
improvement to the thermal feedback prescriptions would make the
results of large-scale numerical simulations much more realistic.

The structure of the paper is as follows. In Section
\ref{sec:nummodel}, we describe the numerical model and setup used in
the simulations. The simulation results are presented in Section
\ref{sec:results}. We discuss the implications of our work for both
numerical modelling of galaxy evolution and interpretation of
observations in Section \ref{sec:discuss} and conclude in Section
\ref{sec:concl}.

\section{Numerical model}\label{sec:nummodel}

\subsection{Basics of the model}

The code used is GADGET-3, an updated version of the publicly
available GADGET-2 \citep{Springel2005MNRAS}. It is a hybrid
N-body/SPH code with individual particle timesteps and adaptive
smoothing. We employ the SPHS extension of the basic SPH scheme
\citep{Read2012MNRAS}, which improves the modelling of mixing within a
multiphase medium, allowing us to better track the interaction between
turbulent gas flows. We use a Wendland kernel \citep{Wendland95,
  Dehnen2012MNRAS} with 100 neighbours.

The initial conditions of all models presented in this paper are the
same. They consist of a gas sphere with isothermal density profile,
extending from $R_{\rm in} = 100$~pc to $R_{\rm out} = 1$~kpc. The gas
sphere is resolved with $10^6$ particles, giving a particle mass of
$m_{\rm SPH} = 3000 \msun$ and a mass resolution of $m_{\rm res} =
100m_{\rm SPH} = 3\times 10^5 \msun$. The spatial resolution of the
model varies depending on gas density, but is generally of order a few
pc; we adopt a minimum gravitational and SPH smoothing length of
$0.1$~pc. The gas is embedded in a static isothermal background
density profile with velocity dispersion $\sigma = 200$~km/s, and is
supported against the background gravity by turbulent motion with
characteristic velocity $v_{\rm turb} = \sigma$. The gas is initially
cold ($c_{\rm s} \ll v_{\rm turb}$; see below for a discussion on how
the gas evolves thermally). A SMBH is embedded in the centre of the
gas distribution and can swallow gas which comes within the accretion
radius $r_{\rm accr} = 10$~pc of the SMBH particle and is
energetically bound to it. We set the SMBH particle mass to $2 \times
10^8 \; \msun$, in order for the radius of its sphere of influence to
be $r_{\rm infl} \simeq r_{\rm accr}$. This ensures that the gas moves
in a purely isothermal background potential only perturbed by its own
self-gravity, rather than the gravity of the SMBH. The SMBH is
inactive for the first $1$~Myr of the simulation; this allows the gas
to develop a turbulent density structure before it is affected by
feedback. In order to prevent the formation of spurious high-density
filaments close to the inactive SMBH, we extend the accretion radius
to $r'_{\rm accr} = R_{\rm in}$ during this time. After $1$~Myr, the
AGN switches on for $1$~Myr and begins affecting the gas in two ways.

First of all, we model the effect of the AGN radiation field by
employing a \citet{Sazonov2005MNRAS} cooling curve, appropriate for
optically thin gas subjected to AGN radiation. Below $T=10^4$~K, we
extend the cooling curve with the one proposed by
\citet{Mashchenko2008Sci} down to a temperature floor, which depends
on the gas density such that the Jeans mass of the gas never falls
below the resolution limit.

Gas particles which fall on this dynamic temperature floor are
stochastically converted into sink particles in order to speed up the
simulations. This is equivalent to having a density threshold
\begin{equation}
\rho_{\rm J} = \left(\frac{\pi k_{\rm B} T}{\mu m_{\rm p} G}\right)^3
m_{\rm res}^{-2} \simeq 6.7 \times 10^{-15} T_4^3 \; {\rm g}
{cm}^{-3},
\end{equation}
\begin{equation}
n_{\rm J} \simeq 6.2 \times 10^{9} T_4^3 \; {\rm cm}^{-3},
\end{equation}
where $T_4 \equiv T/10^4$~K. This crude approximation of the star
formation process may affect our results slightly, because sink
particles are affected only by gravity and cease to interact with the
AGN feedback. Since sink particles are created in dense gas, this
slightly reduces the fraction of feedback energy received by dense gas
and increases the fraction received by diffuse gas. On the other hand,
we do not model the effects of self-shielding of dense gas clumps, so
the reduction in received energy caused by the formation of sink
particles mimics that effect to some extent. In any case, this effect
is small, since our simulations never produce more than $10^4$ sink
particles.

The other way that the AGN affects the gas is by means of wind
feedback, which we describe below.

\subsection{Physics and implementation of AGN feedback}

\begin{table*}
\begin{tabular}{c | c c | c c c c}
Model ID & Feedback model & $\frac{L_{\rm AGN}}{1.3\times10^{46} {\rm erg s}^{-1}}$ & $r_{\rm bub}$ (kpc) & $\frac{E_{\rm gas}}{E_{\rm input}}$ & $\dot{M}_{\rm SMBH}$ ($\msun /$~yr$^{-1}$) & $\dot{M}_{\rm out}$ ($\msun /$~yr$^{-1}$)\\
\hline
\hline
NoAGN     & None & $0$ & $0$ & $-$ & $130$ & $0$ \\
\hline
vp-L1  & Virtual particle & $1$ & $0$ & $0.01$ & $750$ & $40$ \\
vp-L2  & Virtual particle & $2$ & $0.2$ & $0.01$ & $550$ & $200$ \\
vp-L5  & Virtual particle & $5$ & $0.5$ & $0.3$ & $8$ & $1300$ \\
\hline
tk-L1  & Thermal          & $1$ & $0.1?$ & $0.01$ & $800$ & $15$ \\
tk-L2  & Thermal          & $2$ & $0.2$ & $<0$   & $900$ & $20$ \\
tk-L5  & Thermal          & $5$ & $0.3$ & $0.05$ & $0$ & $180$ \\
\hline
vpc-L1  & Virtual particle biconical & $1$ & $0.25$ & $0.01$ & $450$ & $130$ \\
vpc-L2  & Virtual particle biconical & $2$ & $0.35$ & $0.04$ & $150$ & $470$ \\
vpc-L5  & Virtual particle biconical & $5$ & $0.75$ & $0.85$ & $20$  & $1900$ \\
\hline
tkc-L1  & Thermal biconical & $1$ & $0.2$  & $0.02$  & $650$ & $90$  \\
tkc-L2  & Thermal biconical & $2$ & $0.35$ & $0.035$ & $400$ & $220$ \\
tkc-L5  & Thermal biconical & $5$ & $0.6$  & $0.3$   & $170$ & $750$ \\
\hline
\hline

\end{tabular}
\caption{Parameters of the numerical models and most important
  results. The first column shows the model ID. The next two columns
  give the parameters: the type of AGN feedback prescription used and
  the AGN luminosity $L_{\rm AGN}$ in units of Eddington luminosity of
  a $10^8 \msun$ SMBH. Then four columns give the primary results at
  $t = 0.5$~Myr after the AGN switches on: the size of the feedback
  bubble in kpc, the ratio between gas energy and input energy from
  the AGN, the mass accretion rate by the SMBH particle, and the mass
  outflow rate.}
\label{table:param}
\end{table*}

The AGN wind feedback model \citep{King2003ApJ, King2010MNRASa} has a
number of appealing properties: it is based upon relatively
well-understood physical processes; it can explain the $M-\sigma$
relation without requiring free parameters; it can explain the
properties of observed fast AGN winds on sub-parsec scales, as well as
those of massive AGN outflows on kpc scales
\citep{Zubovas2012ApJ}. Observations show that winds with velocity
$v_{\rm w} \sim 0.1 c$ and power $L_{\rm kin} \sim 0.05 L_{\rm AGN}$
are present in a large fraction of AGN \citep{Tombesi2010A&A,
  Tombesi2010ApJ}, suggesting a wide opening angle
\citep{Nardini2015Sci}. In the wind feedback model, these winds shock
against the galactic ISM and drive large-scale outflows
\citep{King2010MNRASa}. The shocked wind passes its energy and
momentum to the ISM, accelerating the outflow to $v_{\rm out}
>1000$~km/s \citep{Zubovas2012ApJ}. However, a large fraction of the
outflow energy can leak out through gaps in the non-uniform ISM,
leading to a situation where some of the gas is outflowing with large
velocities, while dense gas is simultaneously inflowing toward the
SMBH \citep{Nayakshin2014MNRAS, Zubovas2014MNRASb}.

Therefore, it is important to check how well numerical simulations can
capture this complex process of simultaneous inflow and outflow on
scales from tens to thousands of parsecs. We do this by exploring two
subgrid feedback implementations and their improvements.

\subsubsection{Thermal feedback model}

The simplest feedback prescription, often used in cosmological models,
is the ``thermal feedback'' subgrid prescription. In this
prescription, the feedback energy of the AGN, which we take to be
$\epsilon_{\rm f} = 5\%$ of its luminous energy output over the SMBH
timestep, is passed as thermal energy to the SMBH particle
neighbours. We choose 100 neighbours for the SMBH particle, the same
as for gas particle. Testing showed that using fewer neighbours
results in feedback becoming extremely inefficient, as gas particles
can be accreted by the SMBH even after being heated; however, if the
particles were not accreted, it is likely that the feedback would be
more efficient as the temperature would be higher and so the cooling
time would increase. Meanwhile, using more neighbours dilutes the
feedback and results in faster cooling of gas \citep{Bourne2015MNRAS}.

The feedback energy given to each particle is weighted by the SPH
interpolation kernel. This means that gas closer to the SMBH receives
proportionately more energy than gas further away.

In addition to thermal energy, each SPH particle receives a similarly
kernel-weighted fraction of the AGN wind momentum, $L_{\rm AGN} dt/c$,
in a direction radially away from the SMBH. While this is not usually
done in cosmological models, we implement this aspect of feedback in
order to be able to do a more direct comparison with the Monte Carlo
radiative transfer method. Tests showed that the effect of momentum
feedback is negligible compared with the thermal energy input.

\subsubsection{Biconical thermal feedback model}

One issue with the simple thermal feedback prescription is that the
feedback energy is given to the nearest particles without taking their
spatial distribution into account. This implicitly assumes that the
gas distribution is (approximately) spherically symmetric. A real
galaxy contains multiphase gas, with many structures degrading the
spherical symmetry, even if such symmetry might be expected on large
scales. For the AGN, this results in neighbouring gas particles having
very different density and correspondingly very different effective
cross-sections. In principle, feedback affecting each particle should
be attenuated by calculating the optical depth between the SMBH and
this particle. However, doing so is a very time-consuming process,
unfeasible to implement in cosmological simulations, which may have
many AGN at the same time.

Therefore, we propose a simpler method, which we believe captures some
of the important physics. The method relies on injecting the thermal
and kinetic energy in a cone centered on the SMBH instead of
spherically symmetrically. Spherical symmetry is unlikely on the
sub-grid scales on which the wind is launched, since the AGN has a
preferred plane - that of the accretion disc - and winds are likely to
be launched perpendicular to it. So we would expect the wind to be
launched with an axisymmetric geometry. This prescription ensures that
gas in all direction does not experience identical feedback.  Instead,
there is a preferred plane for the gas to accrete in, corresponding to
the plane of the sub-grid accretion disc.

Numerically, we implement this model by calculating a separate
neighbour list for the SMBH particle, looking for neighbours only in a
bicone with a specified opening angle, which is set to 45 degrees in
the simulations presented in this paper. For simplicity, we choose the
axis of the cone to lie in the Z direction. The number of neighbours
in the cone is the same as the number of neighbours used for the
hydrodynamics calculations, and the feedback energy input is weighted
by the value of the SPH interpolation kernel, only using the larger
smoothing length corresponding to the neighbours in the cone.

\subsubsection{Virtual particle model}

The ``virtual particle'' Monte Carlo radiative transfer method
\citep{Nayakshin2009MNRAS} relies on packets of feedback energy, which
are emitted by the AGN as particles and move in straight lines with a
constant velocity, carrying momentum and energy. When the virtual
particle moves into an SPH particle's smoothing kernel, it starts
giving up its momentum and energy over several timesteps, distributing
it proportionately among all SPH particles which contain the virtual
particle in their smoothing kernels. Since the virtual particles are
emitted isotropically, the energy injection is also isotropic,
independent of the distribution of gas around the SMBH. This ensures
that, for example, a dense and small gas clump close to the SMBH does
not receive too much of the feedback energy, because the prescription
self-consistently takes into account the cross section of the
clump. It also means that even gas very far away from the SMBH may
receive a direct energy and momentum input, if the intervening space
is empty, for example having been cleared by a progressing outflow.

We test two versions of the virtual particle model: one with
spherically symmetric emission of tracer particles, and one where the
particles are emitted in a cone with a 45-degree opening angle.

\subsection{Model parameters}

We present the results of 13 models; their identification labels and
main parameters are given in Table \ref{table:param}. One simulation,
NoAGN, is intended as a control and does not have any AGN feedback
(the SMBH is still present and acretes gas). Then four sets of three
models each investigate the effects of different subgrid feedback
prescriptions on feedback from AGN of different luminosities. We
choose $L_{\rm AGN}$ to span a range of observed quasar luminosities;
furthermore, these luminosities encompass the critical luminosity
necessary to drive away gas from the AGN by purely momentum feedback
\citep{King2010MNRASa}. The opening angle of the cone in the thermal
conical (tkc) and virtual particle conical (vpc) simulations is $45$
degrees.

\section{Results}\label{sec:results}

At first, we briefly discuss the properties of the gas distribution
just before the AGN switches on. Then we present the morphology of the
spherically symmetric feedback models, vp and tk, and finally move on
to their conical feedback counterparts, vpc and tkc. Later, we discuss
the resulting gas morphology, outflow energetics, inflow and outflow
rates and the density structure.

\subsection{Initial evolution of the gas shell}

\begin{figure}
  \centering
    \includegraphics[trim = 0 0 4mm 0, clip, width=0.49 \textwidth]{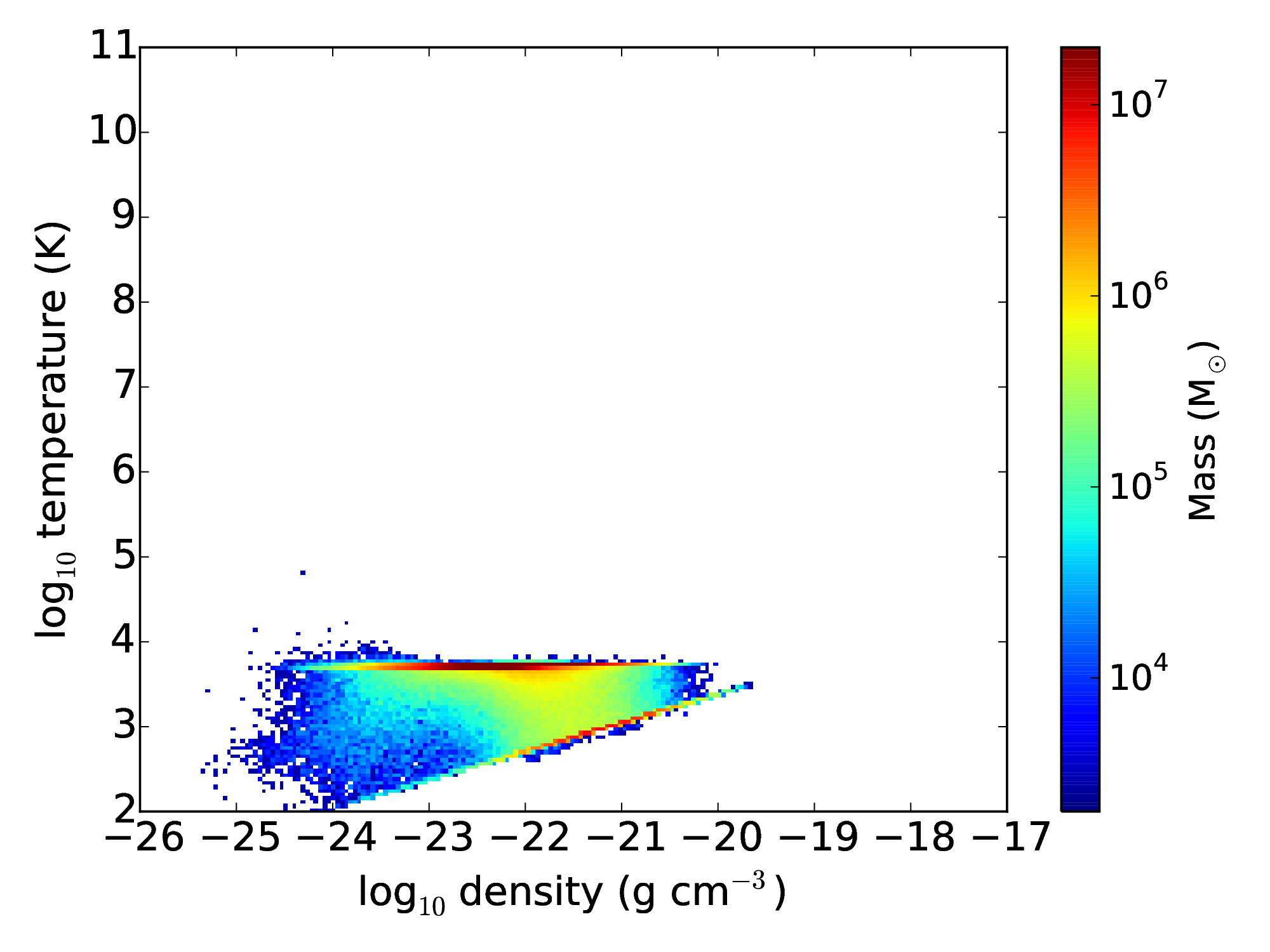}
  \caption{Gas phase diagram just before the AGN switches on. Colour
    indicates relative density of particles in particular regions of
    the diagram, with dark blue being lowest and dark red being
    highest.}
  \label{fig:init_phase}
\end{figure}

During the first 1 Myr of evolution, the gas shell develops a
turbulent density structure and cools down to temperatures in the
range $10 - 10^4$~K. Figure \ref{fig:init_phase} shows the phase
diagram of the gas just before the AGN switches on. Most of the gas
has temperatures just below $10^4$~K, where there is a minimum in the
cooling rate. Some gas is located on the temperature floor (diagonal
line in the bottom right of the distribution), however the
probabilistic conversion of gas into sink particles ensures that only
$\sim2000$ particles have been converted into stars. Approximately
$15\%$, i.e. $4.5\times10^8\msun$, of the initial shell has been
accreted by the SMBH particle. This makes the SMBH mass grow to more
than three times its initial mass, however this does not affect the
subsequent results, since we fix the AGN luminosity, rather than tying
it to the SMBH mass. It is important to note that during this phase,
the accretion radius is set to $100$~pc, and most of the gas accreted
through this radius is unlikely to reach the SMBH during the time
period simulated in our models.

\subsection{Morphology - spherically symmetric feedback}

\begin{figure*}
  \centering
    \includegraphics[trim = 6mm 23mm 6mm 0, clip, width=0.33 \textwidth]{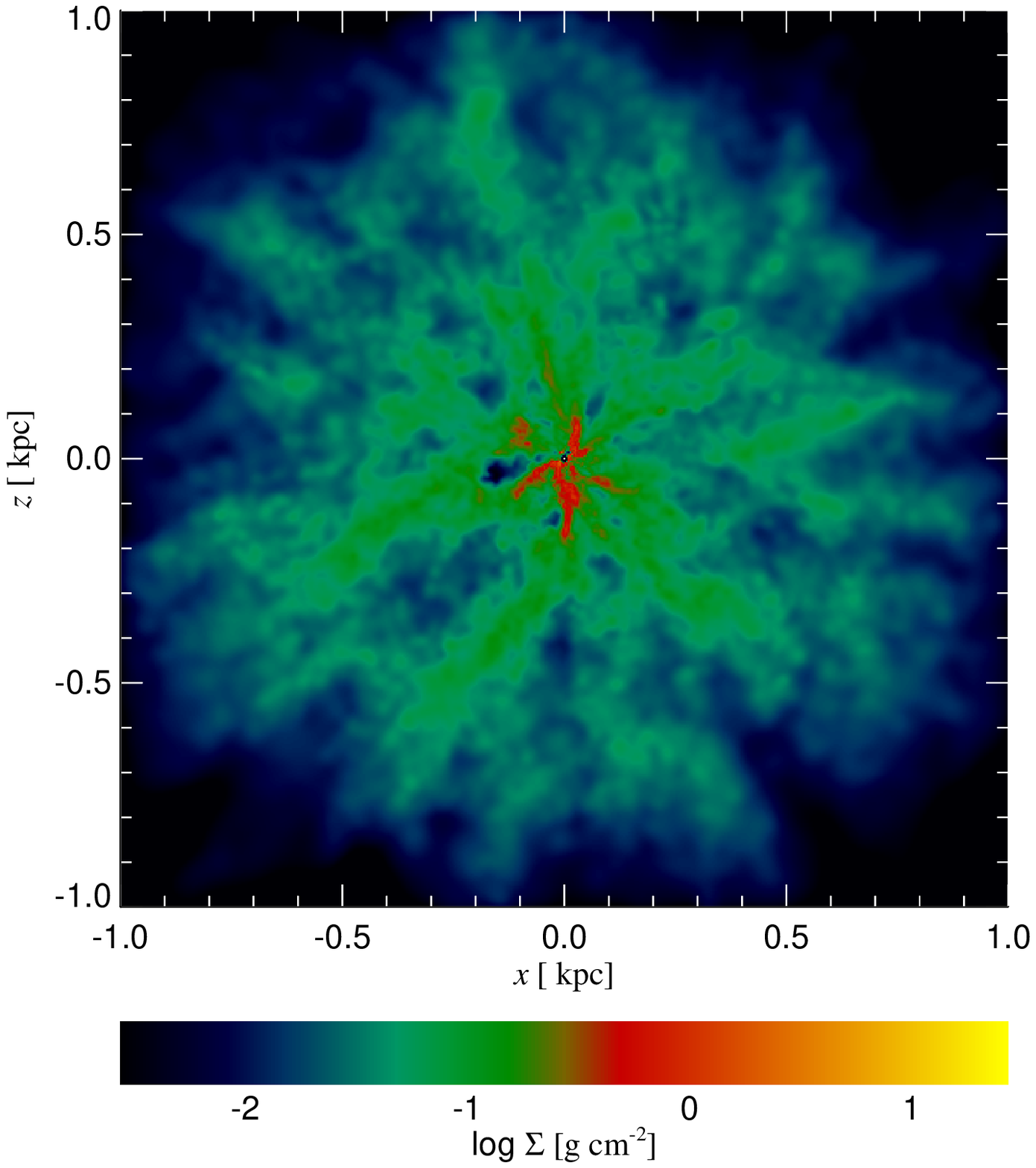}
    \includegraphics[trim = 6mm 23mm 6mm 0, clip, width=0.33 \textwidth]{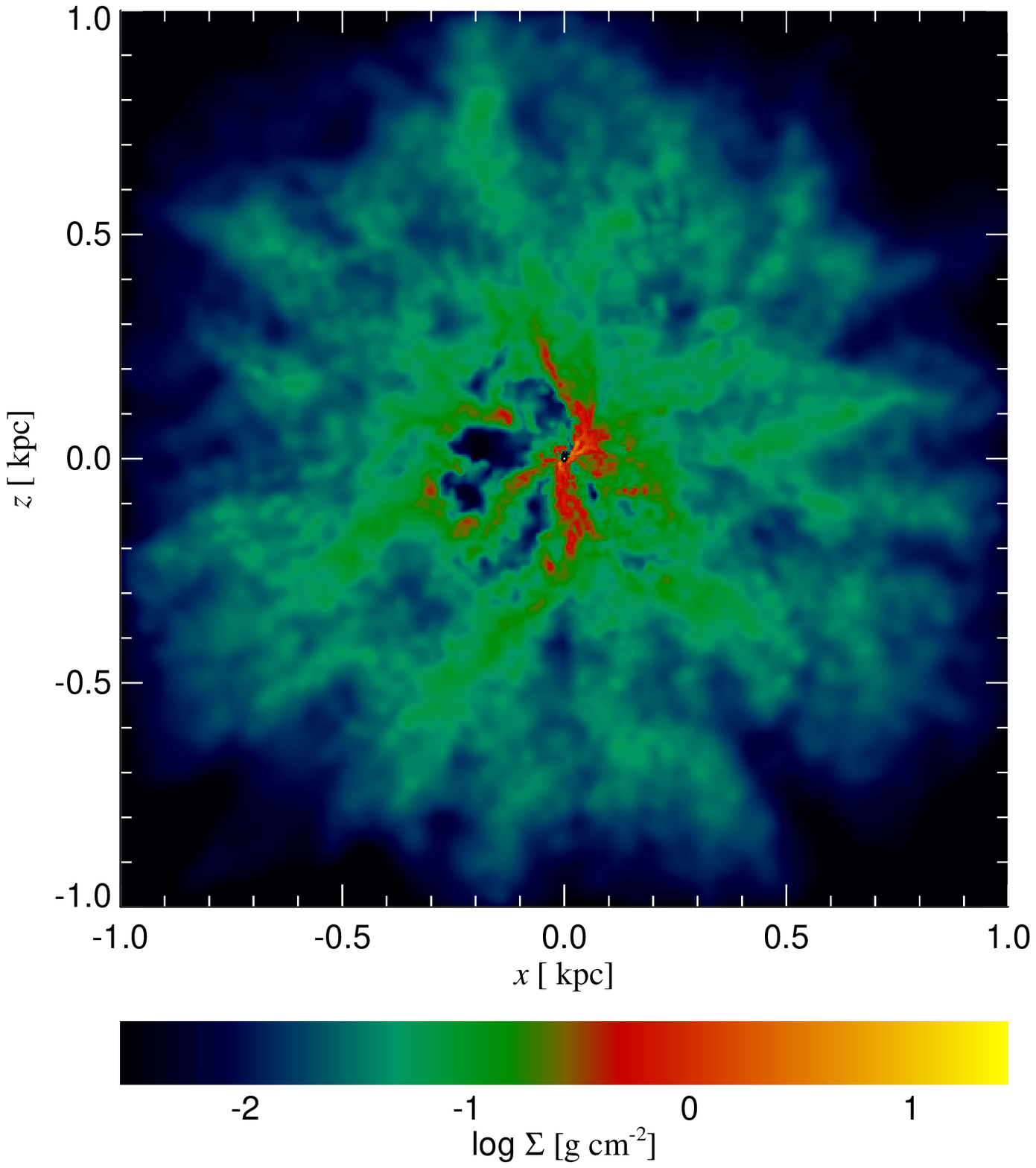}
    \includegraphics[trim = 6mm 23mm 6mm 0, clip, width=0.33 \textwidth]{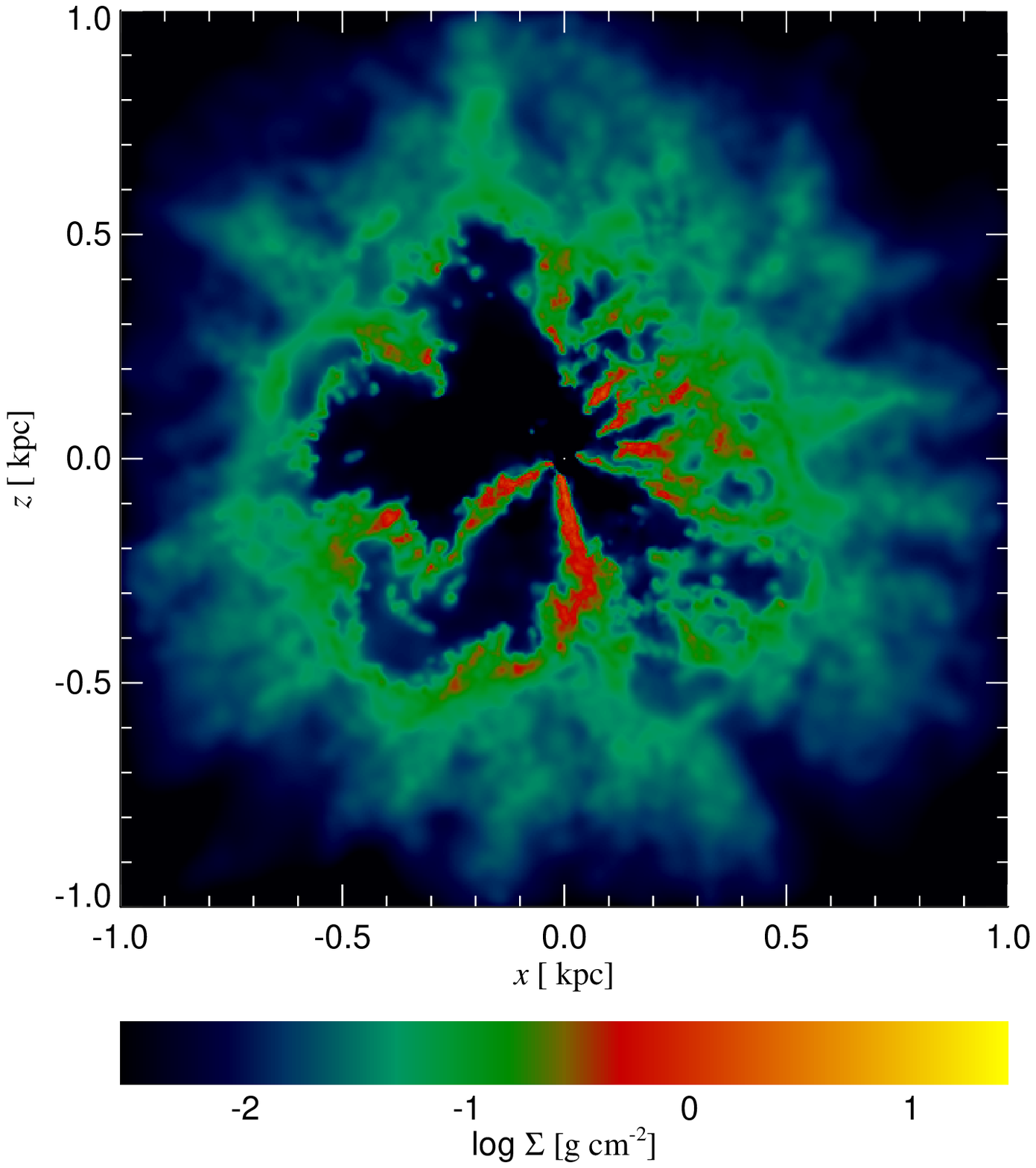}
  \caption{Gas density map in the three virtual particle simulations
    at $t=1.5$~Myr: vp-L1 on the left, vp-L2 in the middle, vp-L5 on the
    right.}
  \label{fig:vp_morph}
\end{figure*}

\begin{figure*}
  \centering
    \includegraphics[trim = 6mm 23mm 6mm 0, clip, width=0.33 \textwidth]{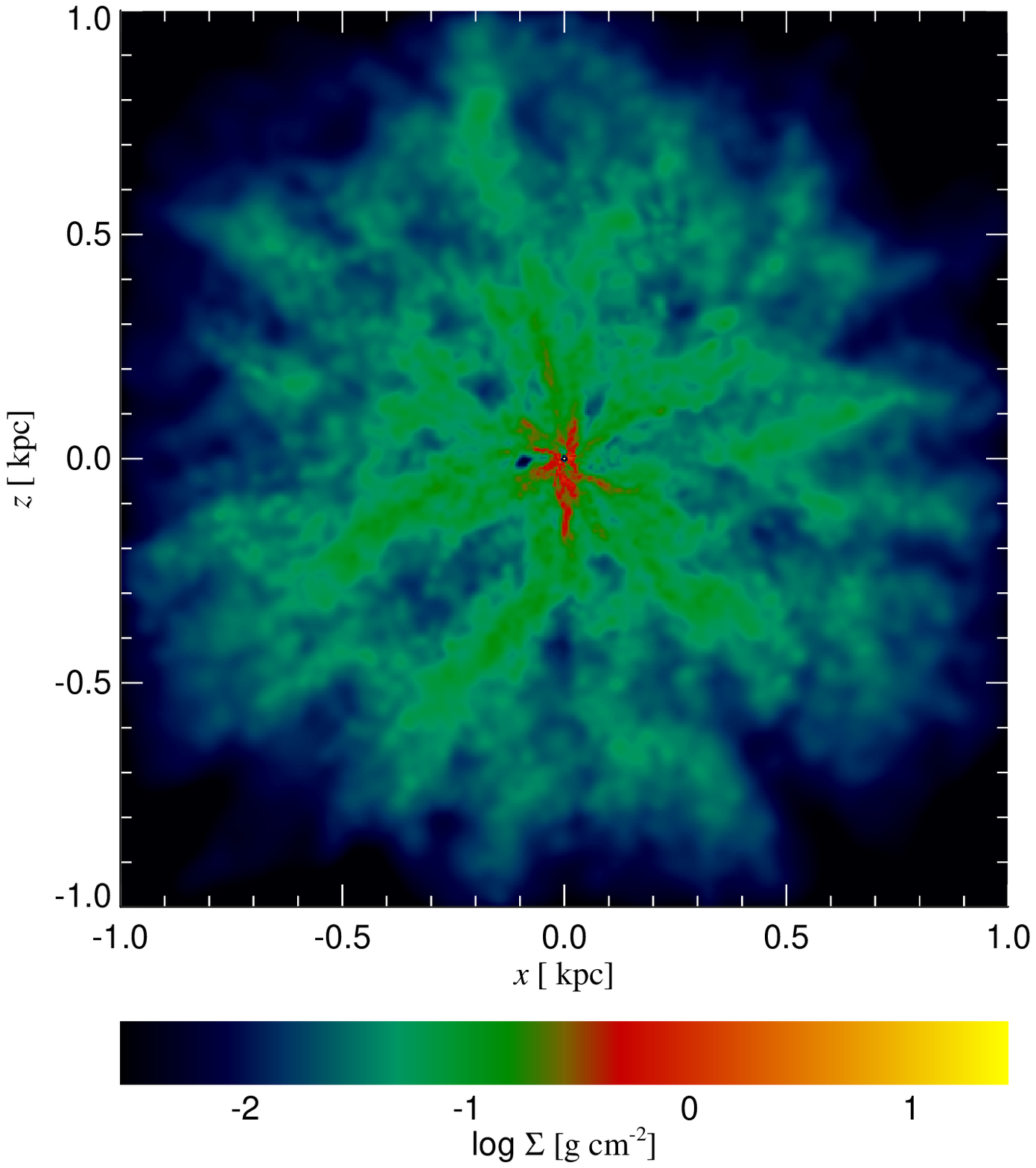}
    \includegraphics[trim = 6mm 23mm 6mm 0, clip, width=0.33 \textwidth]{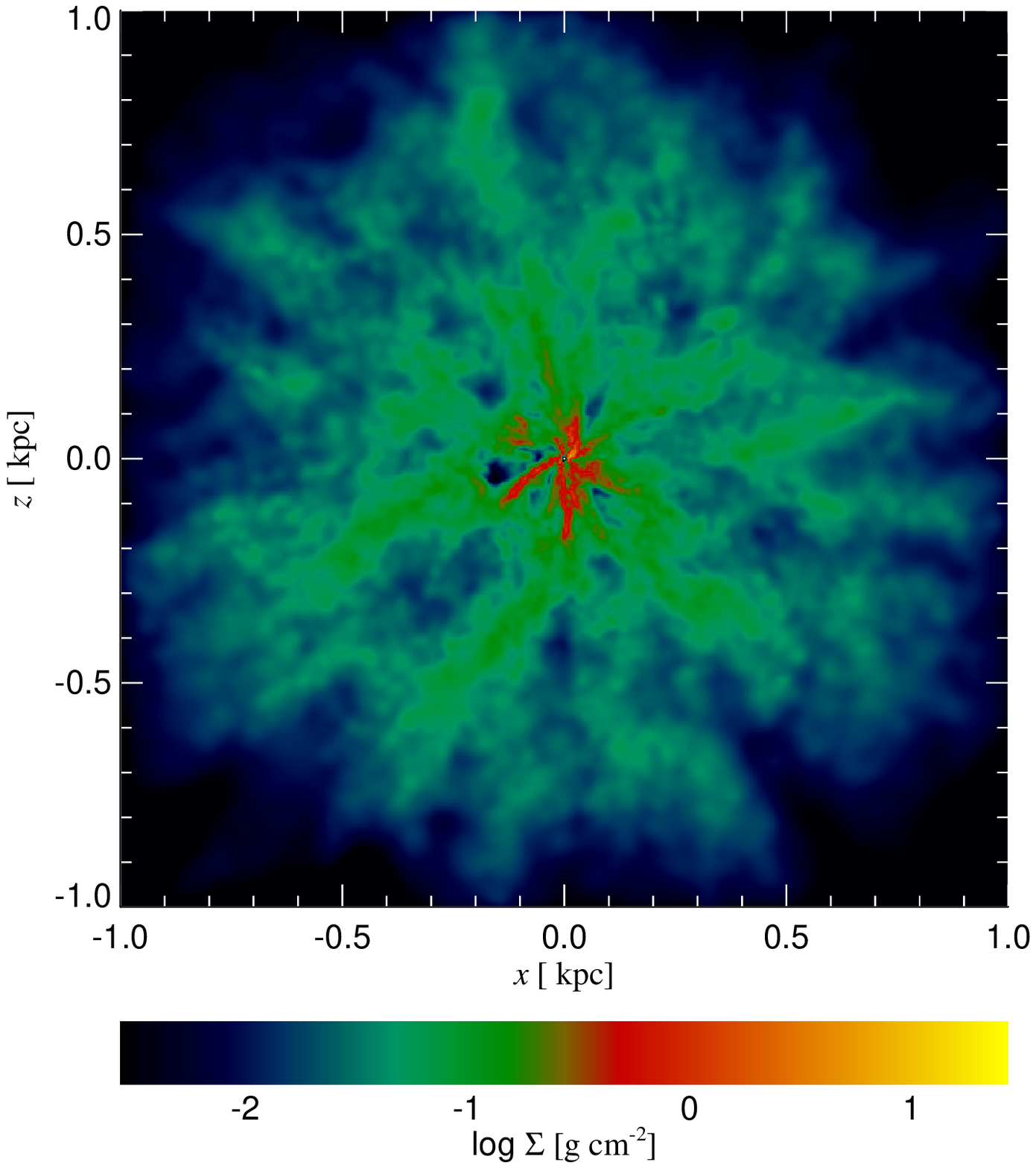}
    \includegraphics[trim = 6mm 23mm 6mm 0, clip, width=0.33 \textwidth]{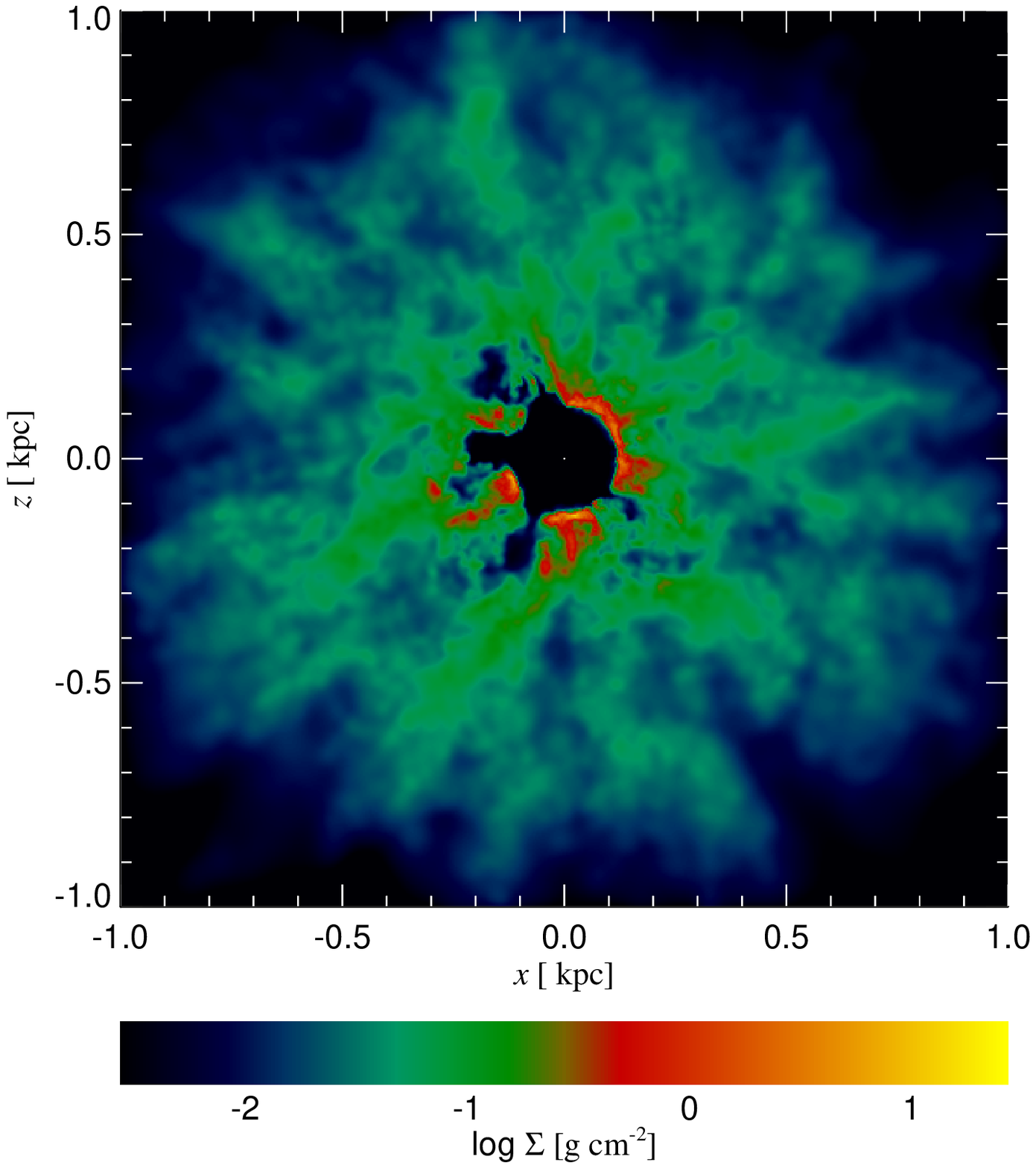}
  \caption{Gas density map in the three thermal models at $t=1.5$~Myr:
    tk-L1 on the left, tk-L2 in the middle, tk-L5 on the right.}
  \label{fig:tk_morph}
\end{figure*}

\begin{figure*}
  \centering
    \includegraphics[trim = 6mm 23mm 6mm 0, clip, width=0.33 \textwidth]{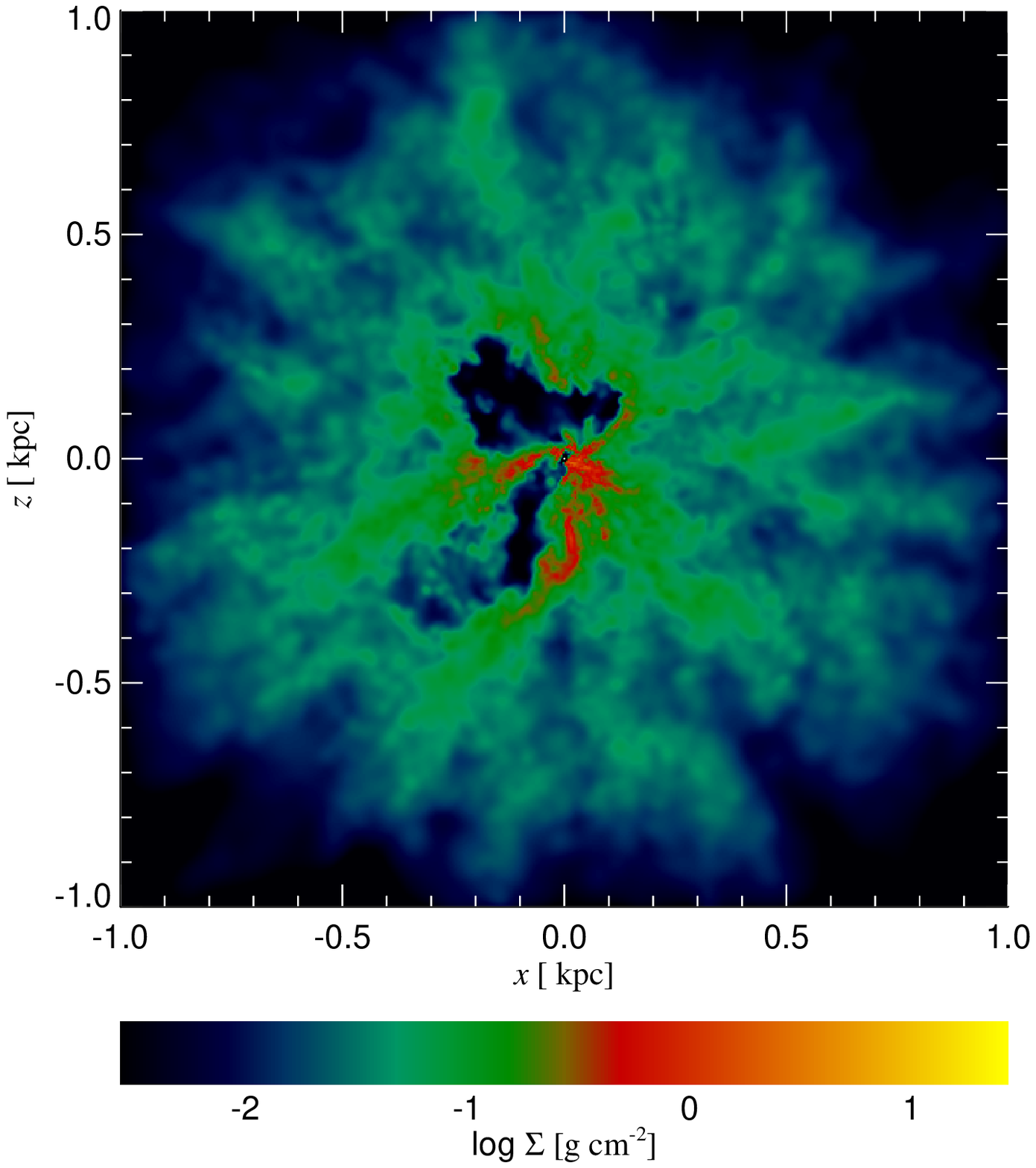}
    \includegraphics[trim = 6mm 23mm 6mm 0, clip, width=0.33 \textwidth]{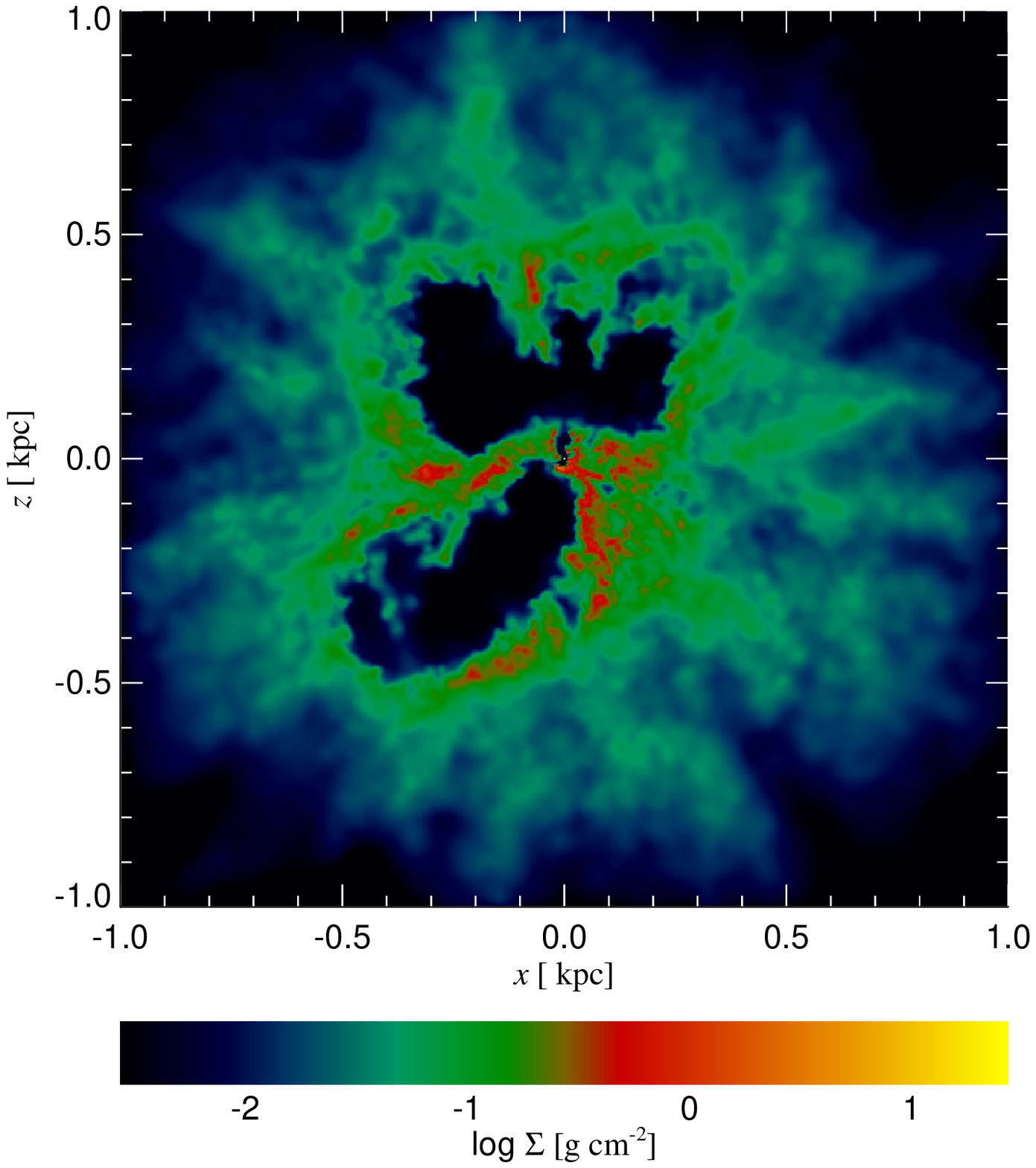}
    \includegraphics[trim = 6mm 23mm 6mm 0, clip, width=0.33 \textwidth]{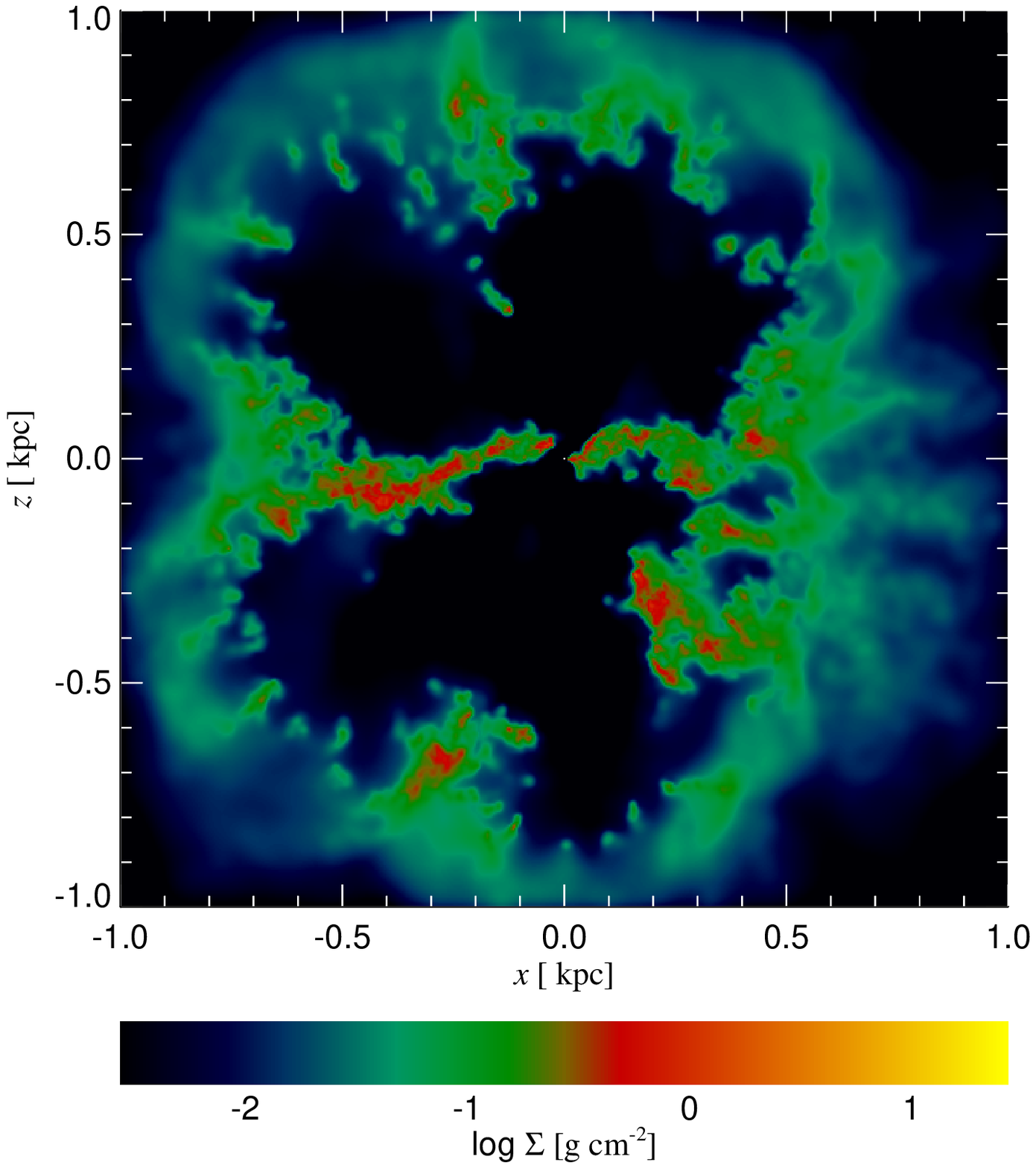}
  \caption{Gas density map in the three conical virtual particle
    simulations at $t=1.5$~Myr: vpc-L1 on the left, vpc-L2 in the
    middle, vpc-L5 on the right.}
  \label{fig:vpc_morph}
\end{figure*}

\begin{figure*}
  \centering
    \includegraphics[trim = 6mm 23mm 6mm 0, clip, width=0.33 \textwidth]{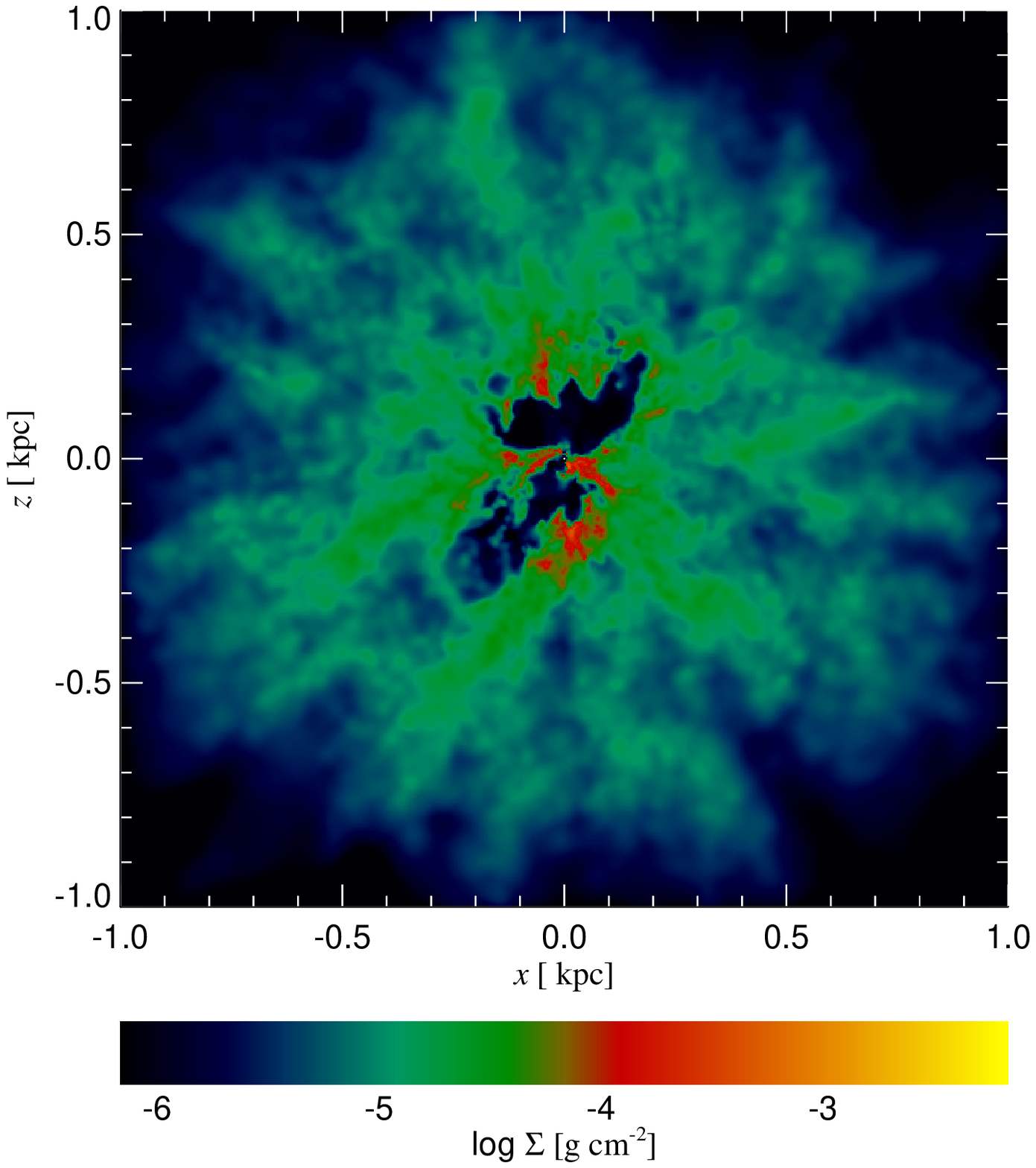}
    \includegraphics[trim = 6mm 23mm 6mm 0, clip, width=0.33 \textwidth]{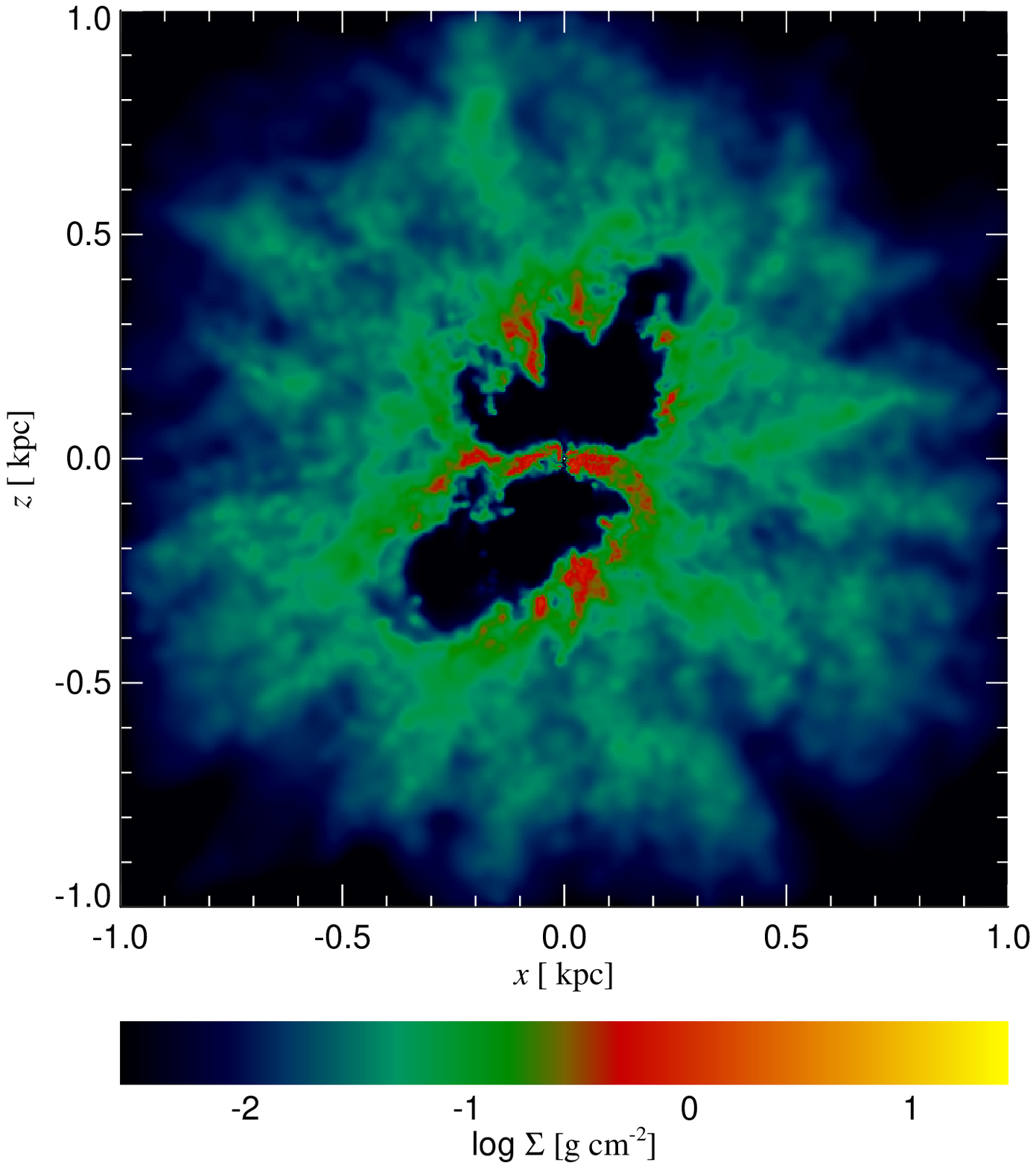}
    \includegraphics[trim = 6mm 23mm 6mm 0, clip, width=0.33 \textwidth]{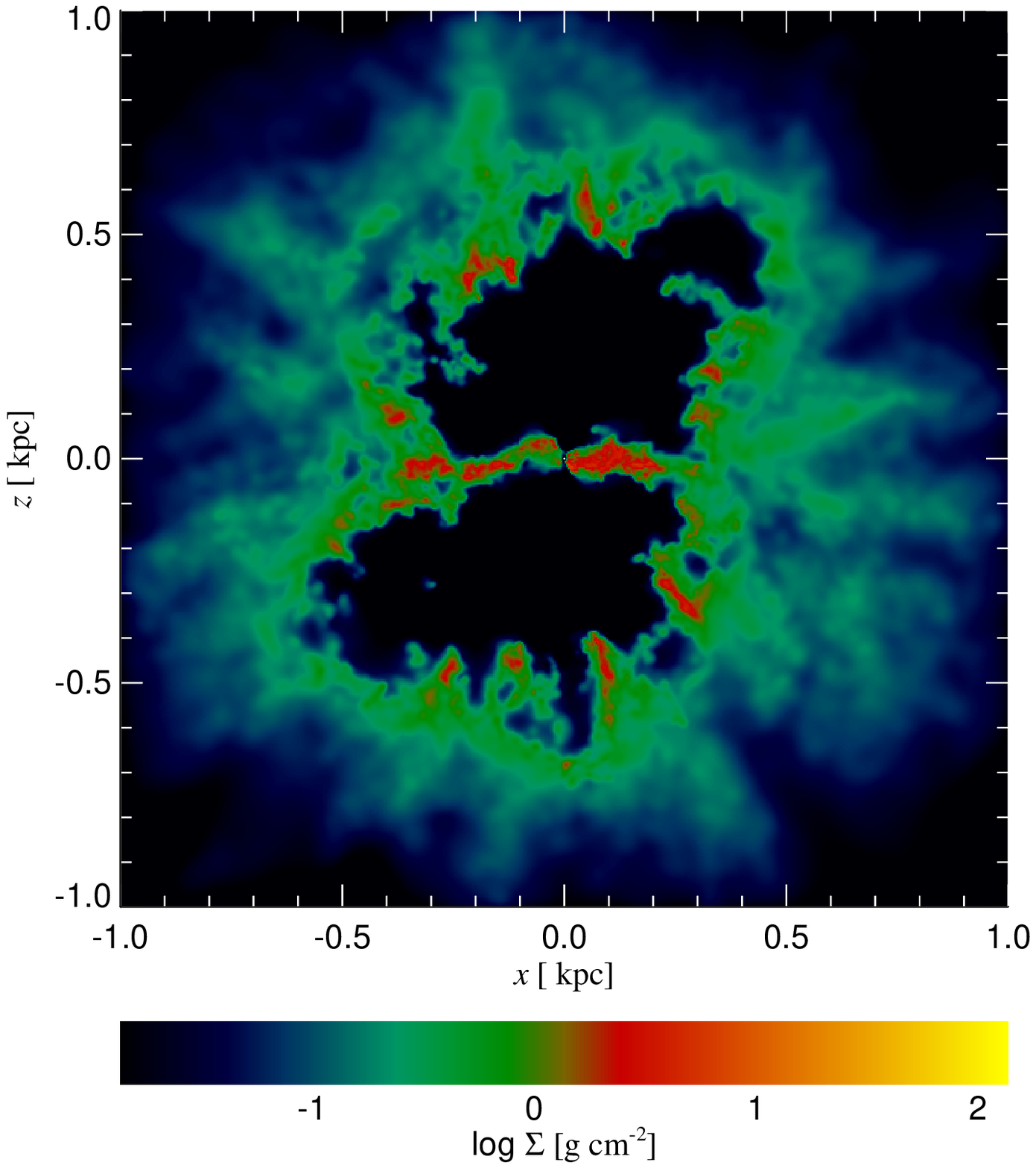}
  \caption{Gas density map in the three conical thermal feedback
    simulations at $t=1.5$~Myr: vpc-L1 on the left, vpc-L2 in the
    middle, vpc-L5 on the right.}
  \label{fig:tkc_morph}
\end{figure*}

Figure \ref{fig:vp_morph} shows the density maps of the three virtual
particle simulations at $t=1.5$~Myr, i.e. $0.5$~Myr after the AGN
switches on, showing the XZ plane. The plots use a wedge-slice
projection, such that only particles with $|y/r| < 0.25$ are plotted,
to ensure that both the innermost regions and the outskirts are
well-sampled. The three plots reveal a clear change in the global
effect of feedback. In the vp-L1 model, feedback is not strong enough
to prevent inflow or even to produce significant bubbles; there is
only one small cavity in the negative-X direction. Simulation vp-L2
shows noticeable outflow bubbles, but simultaneously there exists gas
very close to the SMBH, in fact accreting upon the SMBH particle. Most
of the feedback energy leaks out in the directions of low density,
producing these large bubbles, while the remaining momentum push is
too weak to shut off accretion of dense gas. Finally, in vp-L5, we see
that feedback is powerful enough to blow away almost all the material,
except for a couple of very dense filaments; these remain close to the
SMBH, but are also pushed away, albeit slowly.

In Figure \ref{fig:tk_morph}, we plot density maps of the analogous tk
models, again at at $t=1.5$~Myr. There are significant differences
from the vp models. Most noticeably, a well-defined almost spherical
feedback bubble is present in simulation tk-L5, without any
inflow toward the SMBH. On the other hand, the size of the outflow
bubble is much smaller in tk-L5 than in vp-L5, since the energy is
injected into the gas closest to the SMBH; this gas is dense and
therefore cannot be removed as efficiently as the diffuse gas which
absorbs most of the input energy in the vp models. It is also
interesting to note that the feedback bubble is surrounded by a
boundary layer of dense gas, which is not present in the virtual
particle simulations. This dense gas cools down very rapidly, so most
of the feedback energy is radiated away rather than used to drive
bubble expansion.

The major difference between the two models arises because in the
thermal feedback simulation, all of the feedback energy is always
injected into gas that is closest to the SMBH, typically within the
central $<200$~pc, although the precise radius depends on AGN
luminosity and the size of the forming bubble. In the virtual particle
simulations, on the other hand, feedback evacuates cavities which
allow the virtual particles to stream to large distances of order
$500-700$~pc in the highest-luminosity simulation. We now present the
effects of changing feedback geometry to conical, which ensures that
there is always a channel - the midplane of the gas distribution -
through which gas can reach the SMBH, and similarly that there is a
channel perpendicular to the first through which feedback can be
injected at distances comparable to those in the spherical virtual
particle model.

\subsection{Morphology - conical feedback}

\begin{figure*}
  \centering
    \includegraphics[trim = 0 0 4mm 0, clip, width=0.49 \textwidth]{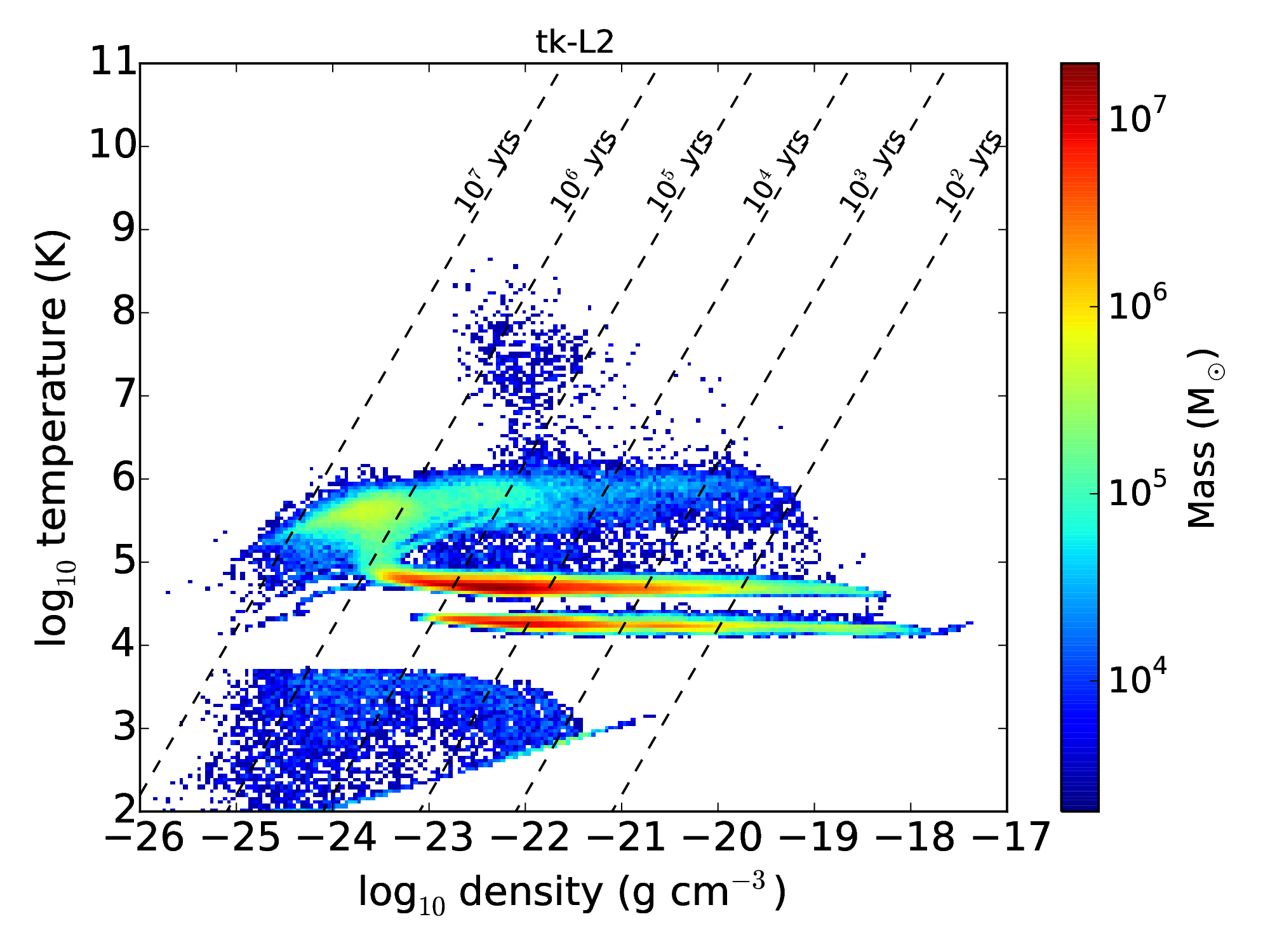}
    \includegraphics[trim = 0 0 4mm 0, clip, width=0.49 \textwidth]{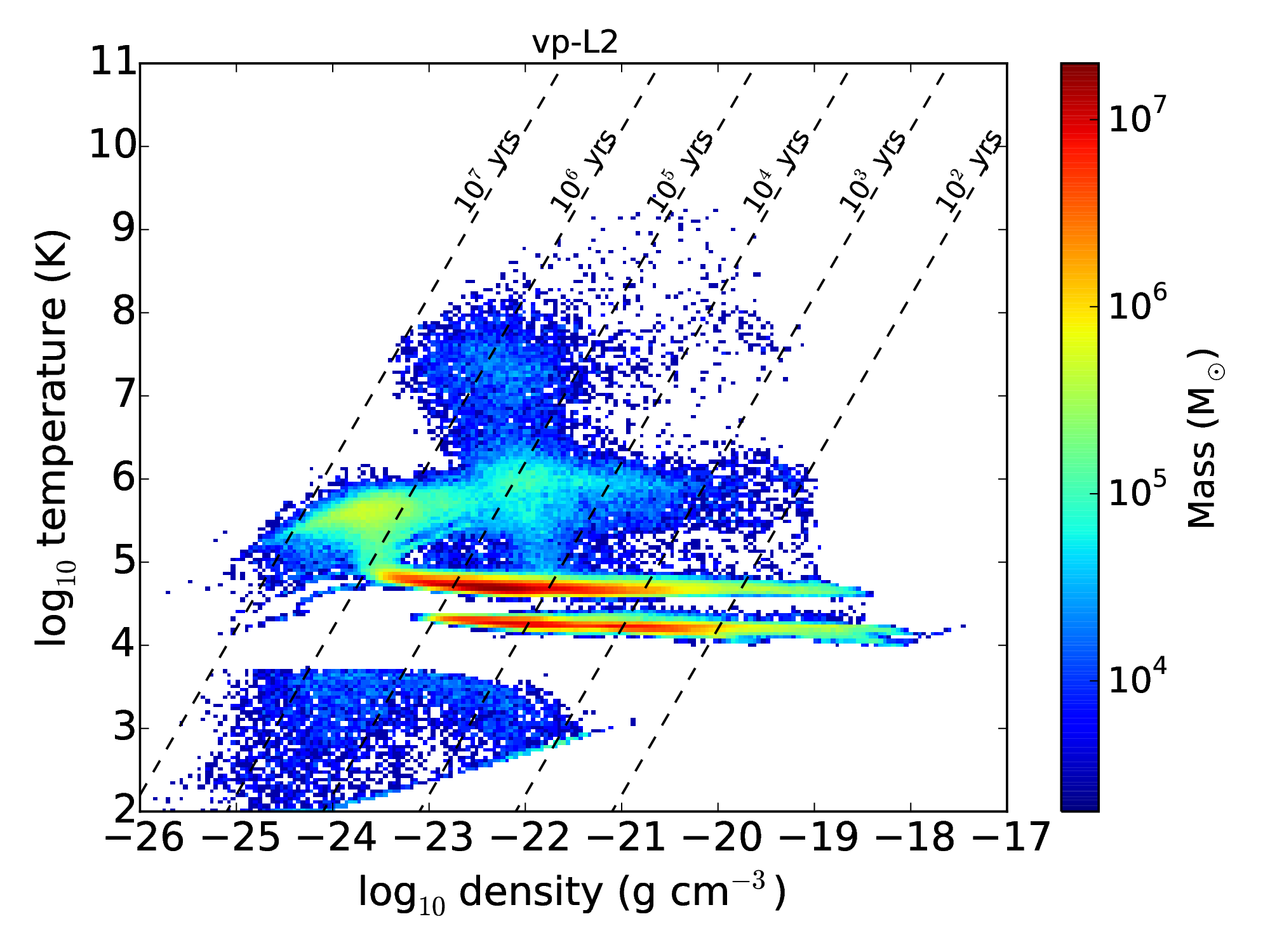}
    \includegraphics[trim = 0 0 4mm 0, clip, width=0.49 \textwidth]{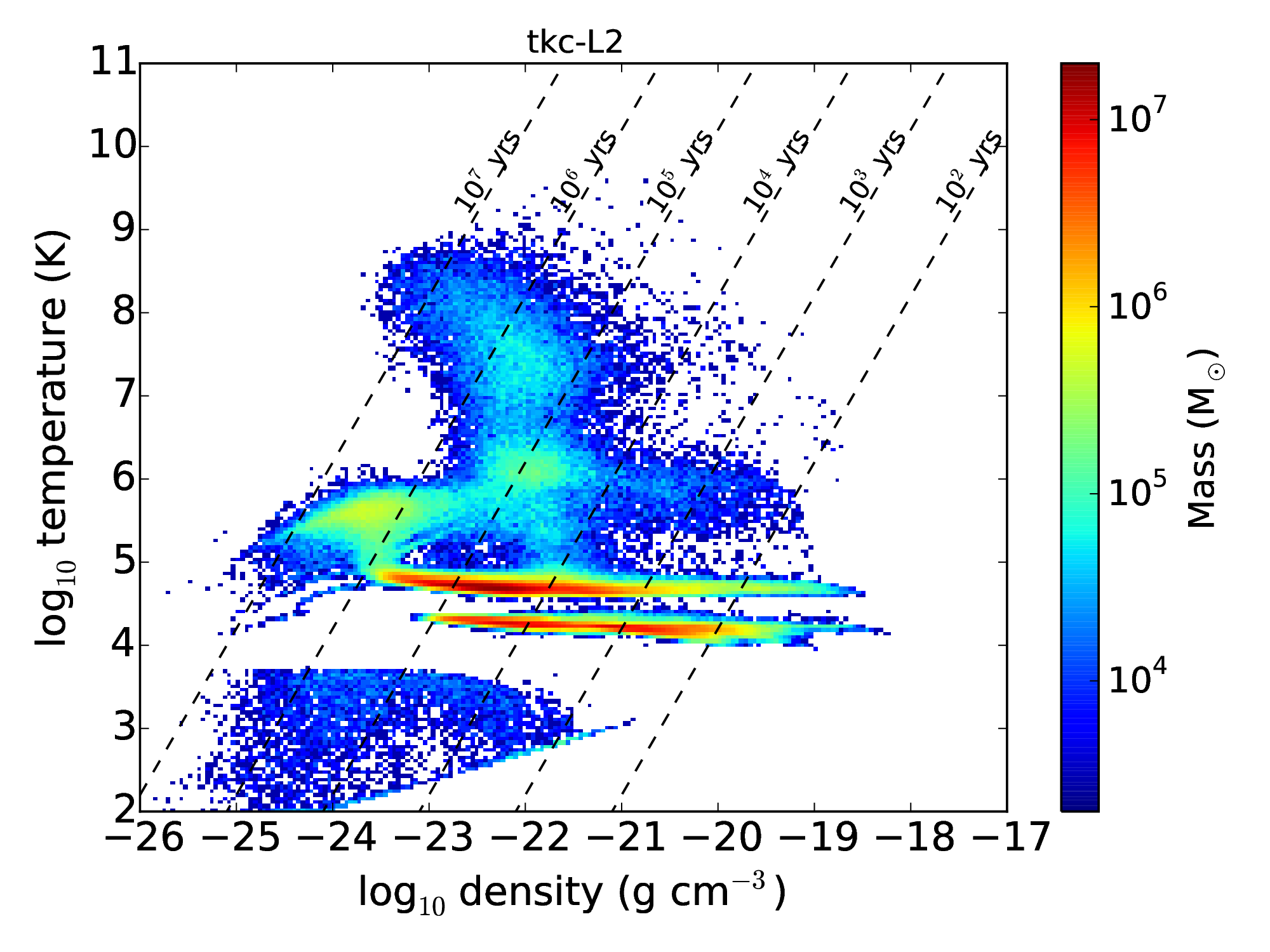}
    \includegraphics[trim = 0 0 4mm 0, clip, width=0.49 \textwidth]{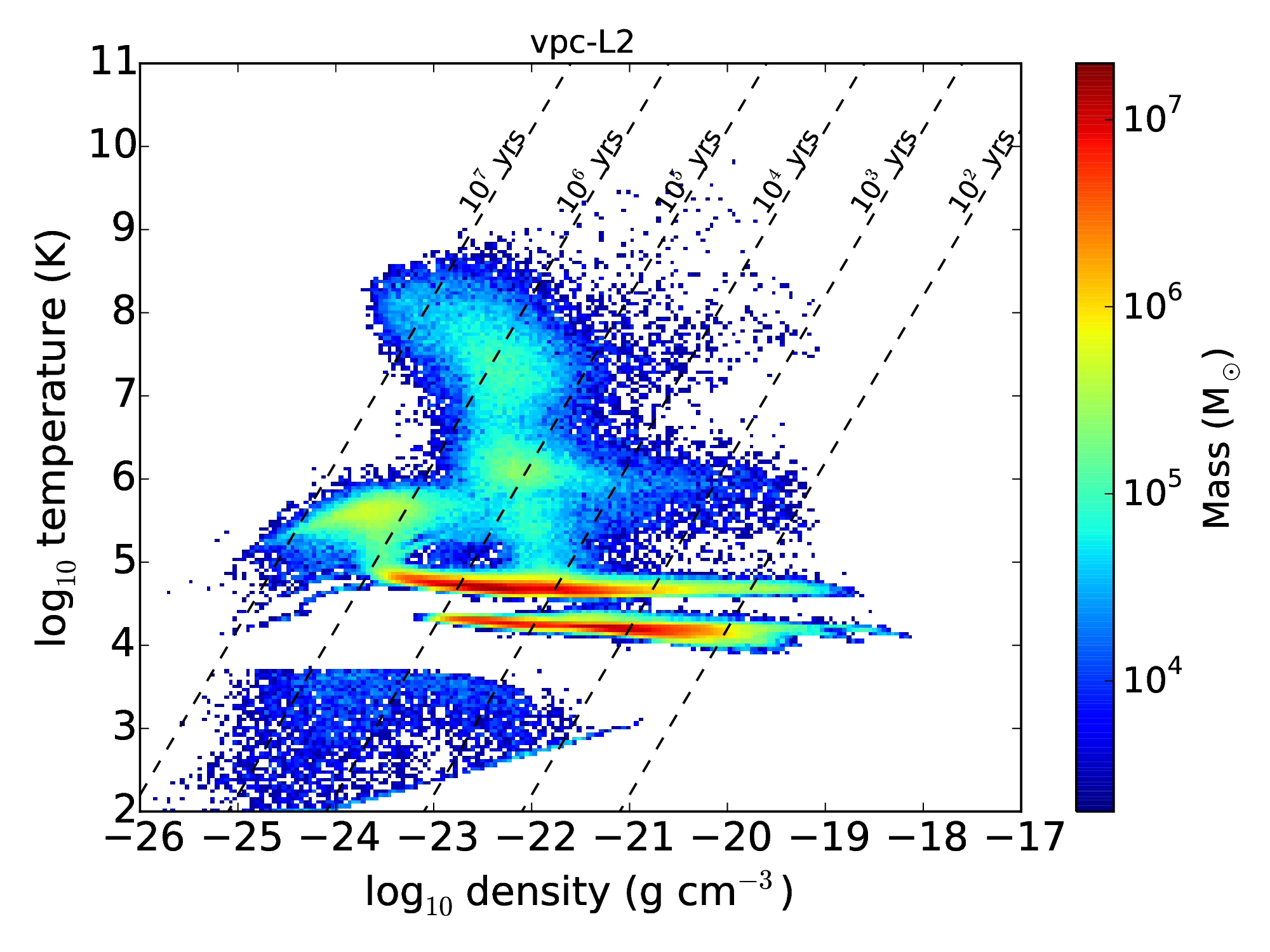}
  \caption{Particle density-temperature plots at t = 1.5 Myr in the
    four L2 simulations: tk-L2 in the top left, tkc-L2 in the top
    right, vp-L2 in the bottom left and vpc-L2 in the bottom
    right. Dashed lines show the expected bremsstrahlung cooling time
    of the gas and colours indicate relative point density, with red
    highest and dark blue lowest.}
  \label{fig:phase_L2}
\end{figure*}

\begin{figure*}
  \centering
    \includegraphics[trim = 0 0 4mm 0, clip, width=0.49 \textwidth]{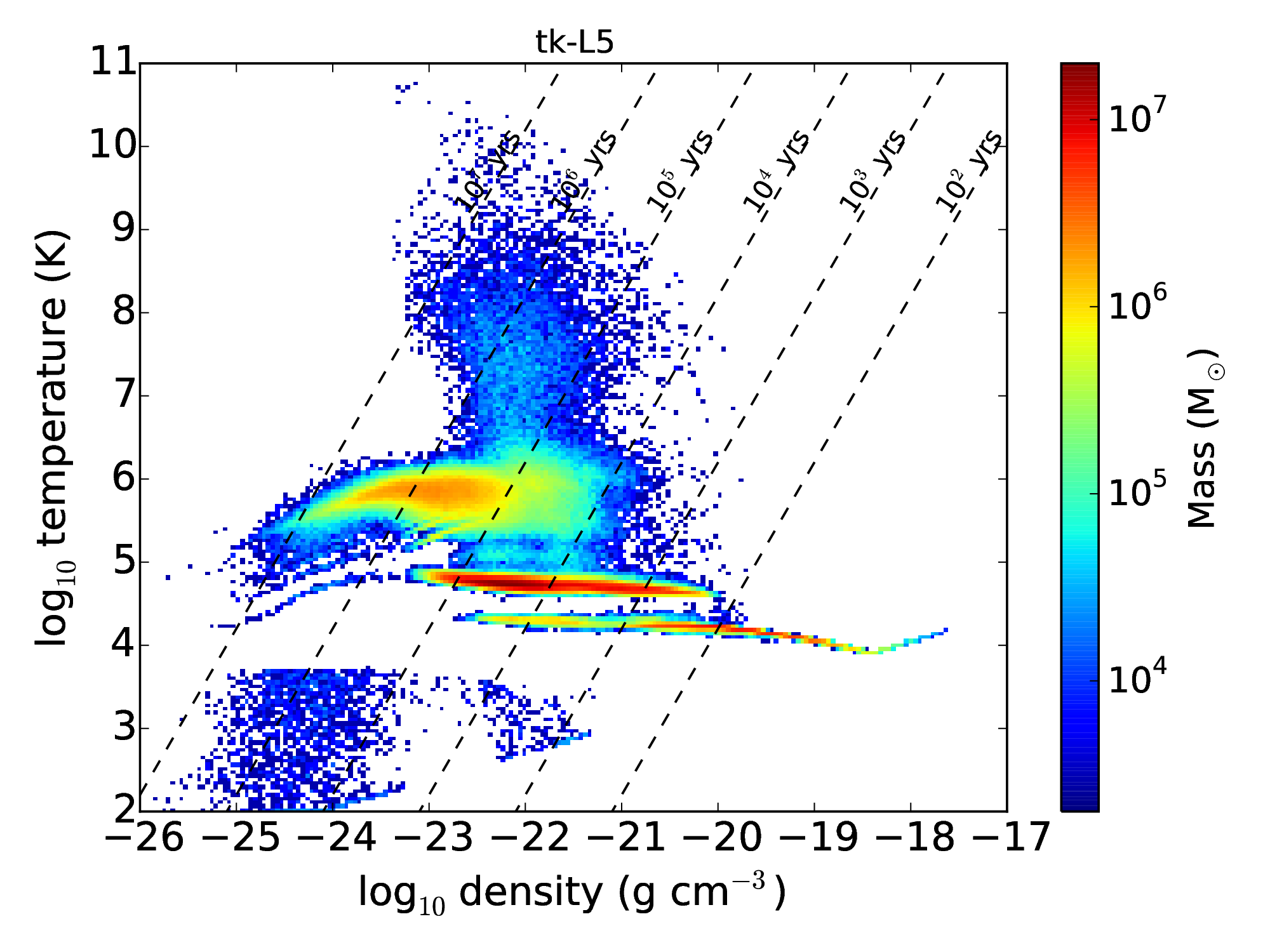}
    \includegraphics[trim = 0 0 4mm 0, clip, width=0.49 \textwidth]{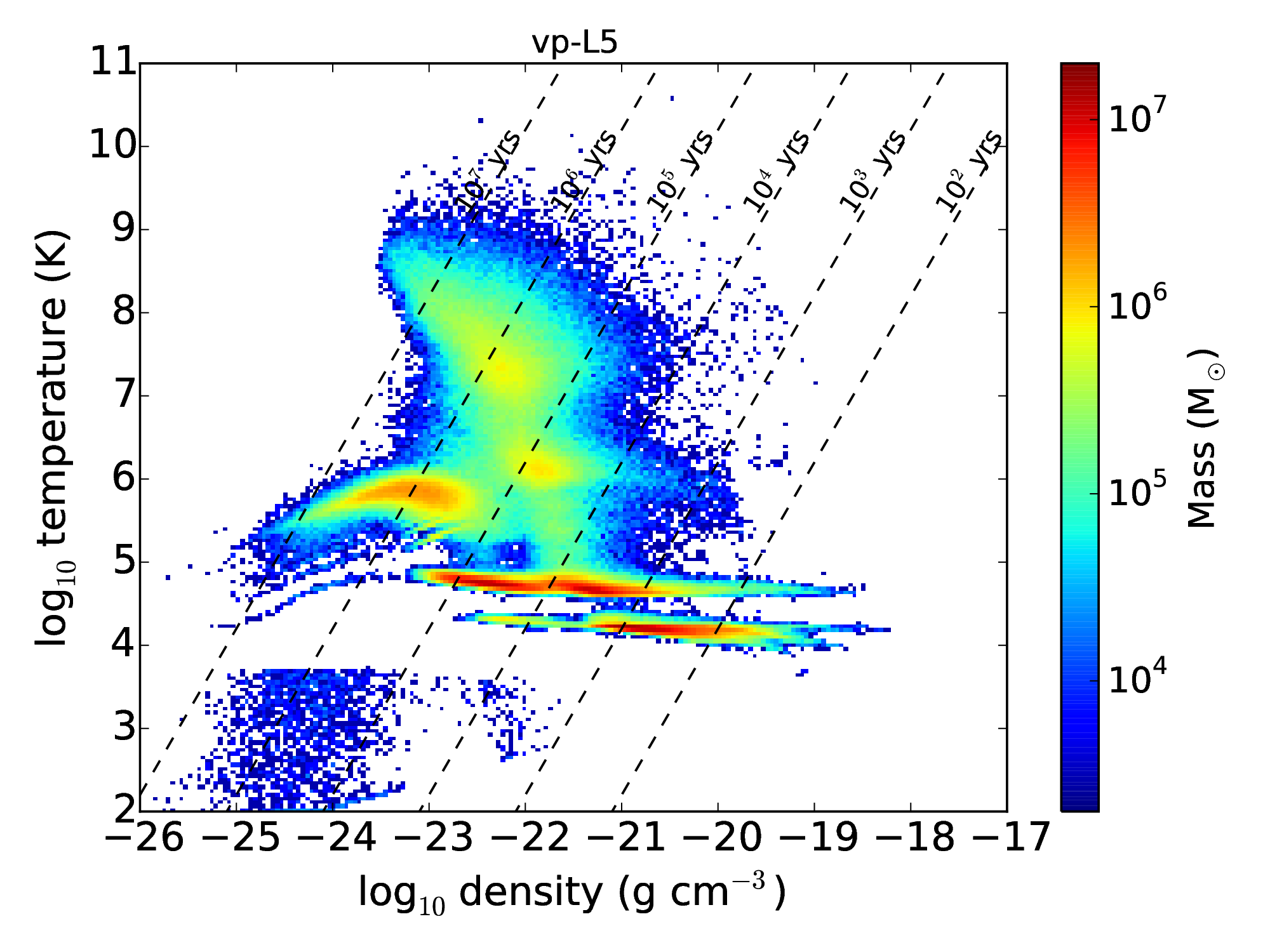}
    \includegraphics[trim = 0 0 4mm 0, clip, width=0.49 \textwidth]{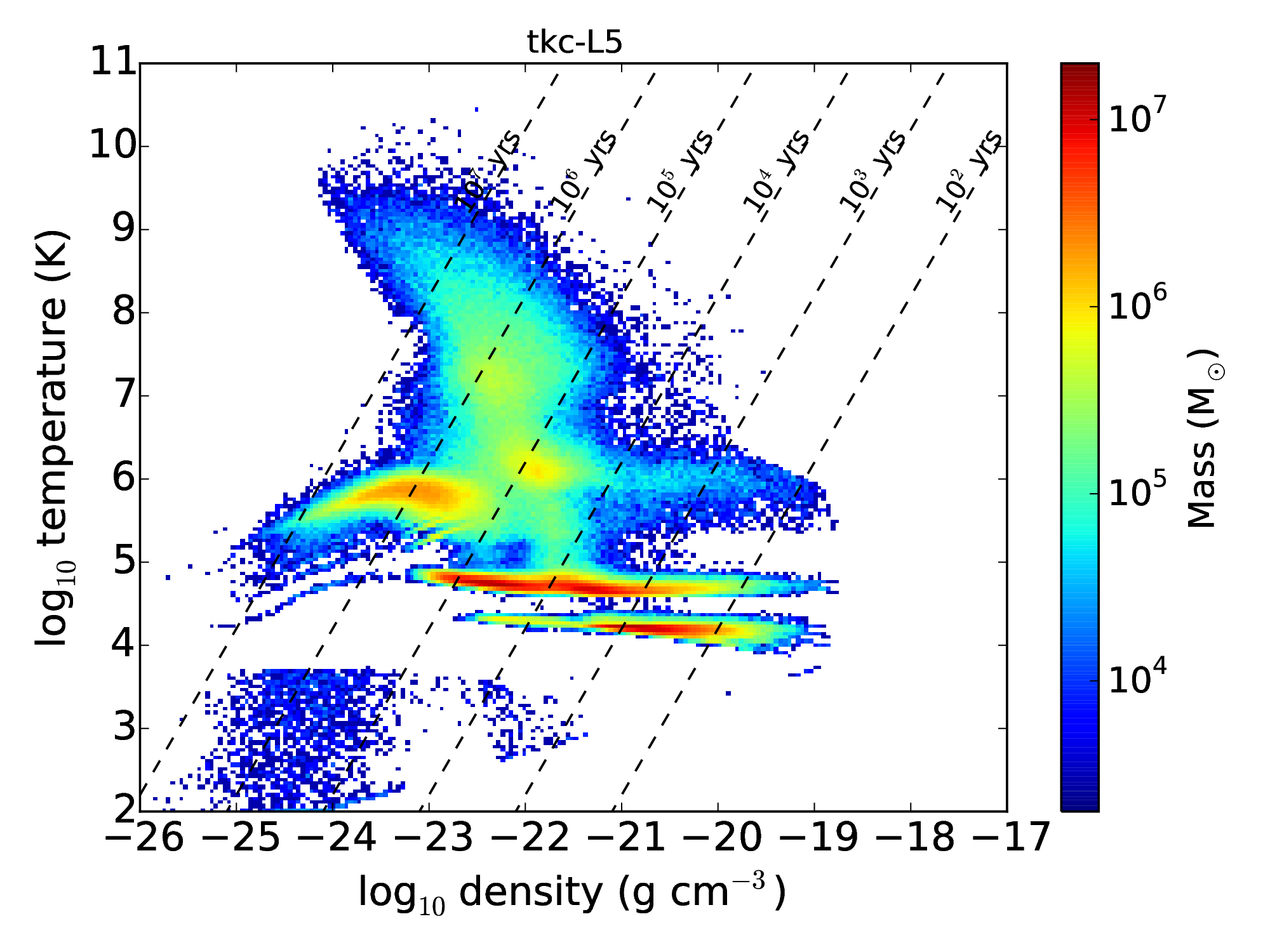}
    \includegraphics[trim = 0 0 4mm 0, clip, width=0.49 \textwidth]{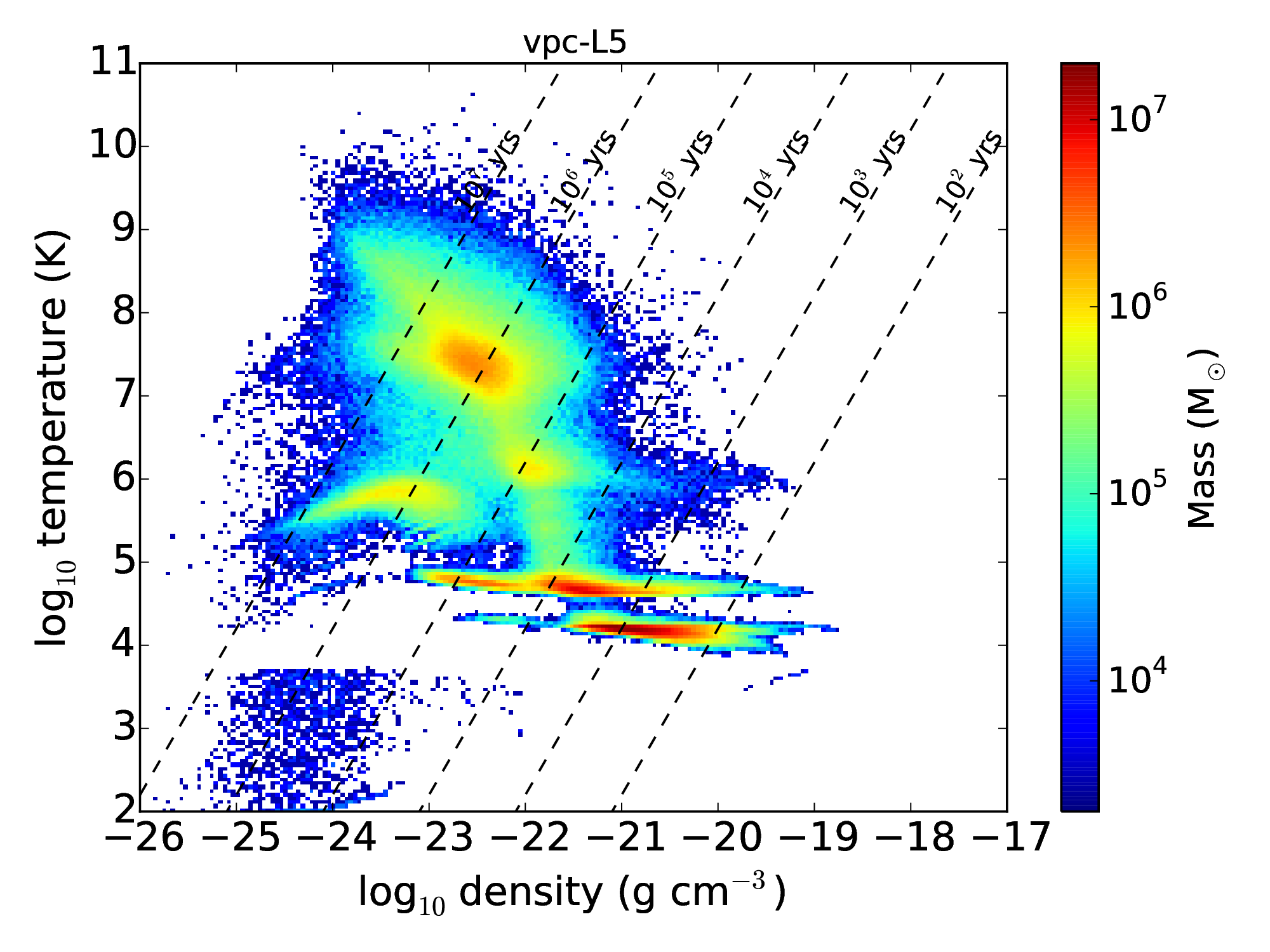}
  \caption{Same as figure \ref{fig:phase_L2}, but for the four L5 simulations.}
  \label{fig:phase_L5}
\end{figure*}

Figure \ref{fig:vpc_morph} shows the gas density in the three conical
virtual particle models, vpc-L1 (left), vpc-L2 (middle) and vpc-L5
(right). Even the lowest-luminosity simulation produces a modest
outflow, with cavities in the top half and third quadrant of the plot,
where the gas was least dense at the time the AGN switched on. The
outflow is somewhat conical, but variations in gas density prevent it
from acquiring a well-defined shape. In the highest-luminosity model,
the feedback bubbles are rather large, touching the edges of the
initial gas distribution, but gas close to the midplane has not been
blown away, allowing filamentary inflow to continue. It is worth
noting that the outflow is wider than the opening angle of the cone
within which the virtual particles are emitted. This happens because
the low-density cavities are over-pressurized and expand both
vertically and laterally.

Figure \ref{fig:tkc_morph} shows the analogous snapshots of the tkc
simulations tkc-L1 (left), tkc-L2 (middle) and tkc-L5 (right). Broadly
speaking, the simulations evolve similarly to the vpc models; there
are two cavities forming in opposite directions and expanding, with
material falling in along the equatorial plane. Overall, the
differences are much smaller than between the tk and vp models
presented above. This happens because with the non-spherical injection
of feedback, a channel appears along the equatorial plane which allows
inflow and outflow to coexist. Therefore, dense gas no longer
accumulates at the edge of a well-defined outflow bubble, but is
brought to the midplane by the same processes that act in the vpc
models. A similar structure of dense inflow along the midplane and
diffuse outflow in the polar direction develops, with some
irregularities due to the turbulent gas density distribution. In the
L1 and L2 simulations, both tkc and vpc models produce similarly-sized
bubbles, while the brightest AGN produces a significantly larger
bubble in the vpc simulation. This last size difference is exacerbated
by the fact that the bubble reaches the edge of the initial gas
distribution in the vpc simulation, and its expansion accelerates. At
earlier times, the difference in size between tkc-L5 and vpc-L5
bubbles is much less pronounced.

Despite the morphological differences and similarities, we are more
interested in the effect that AGN feedback would have upon the
evolution of the host galaxy. Therefore, we now consider several
integrated parameters of the system and their time evolution.

\subsection{Energetics}

\begin{figure}
  \centering
    \includegraphics[trim = 0 0 4mm 0, clip, width=0.49 \textwidth]{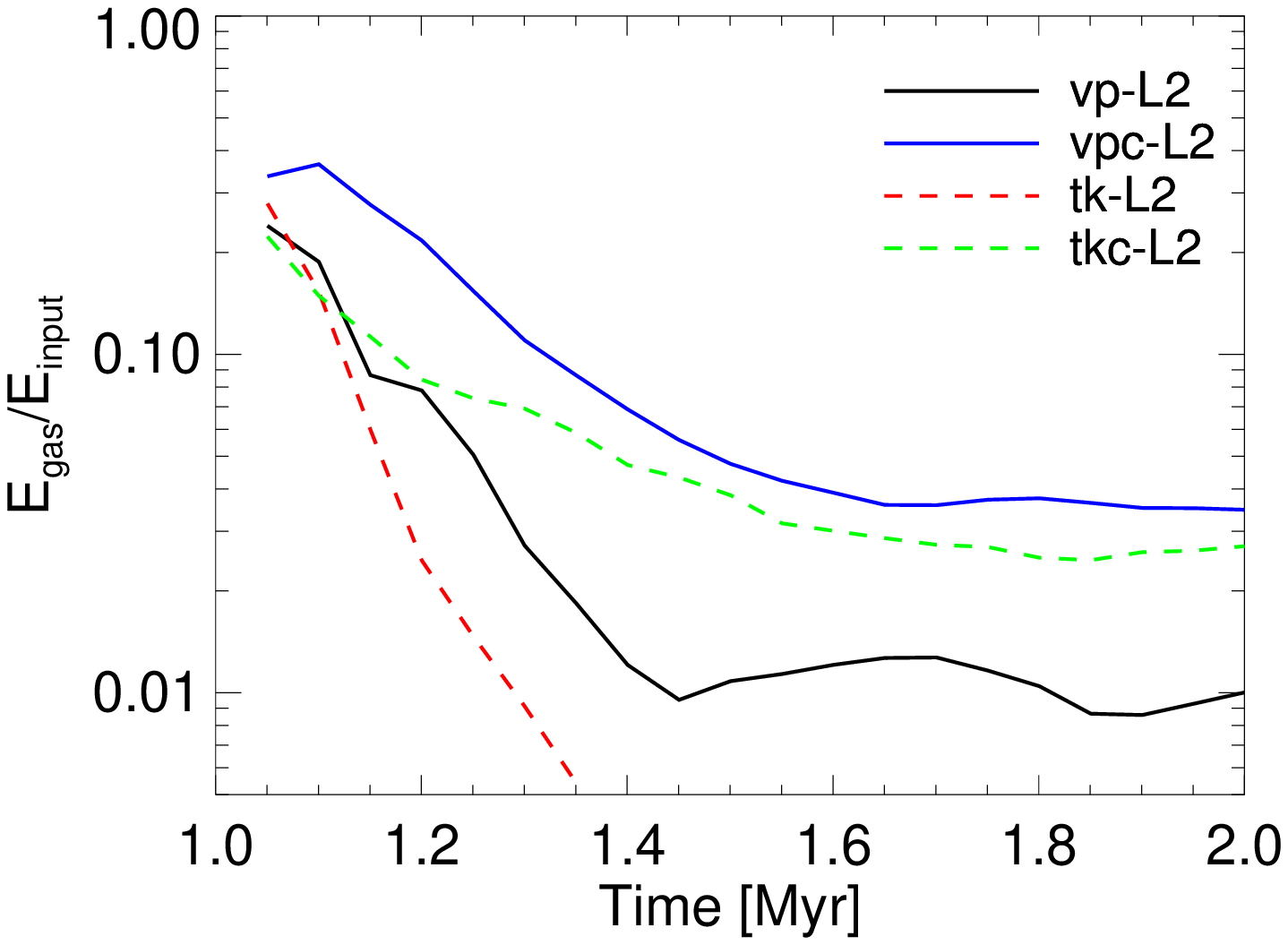}
    \includegraphics[trim = 0 0 4mm 0, clip, width=0.49 \textwidth]{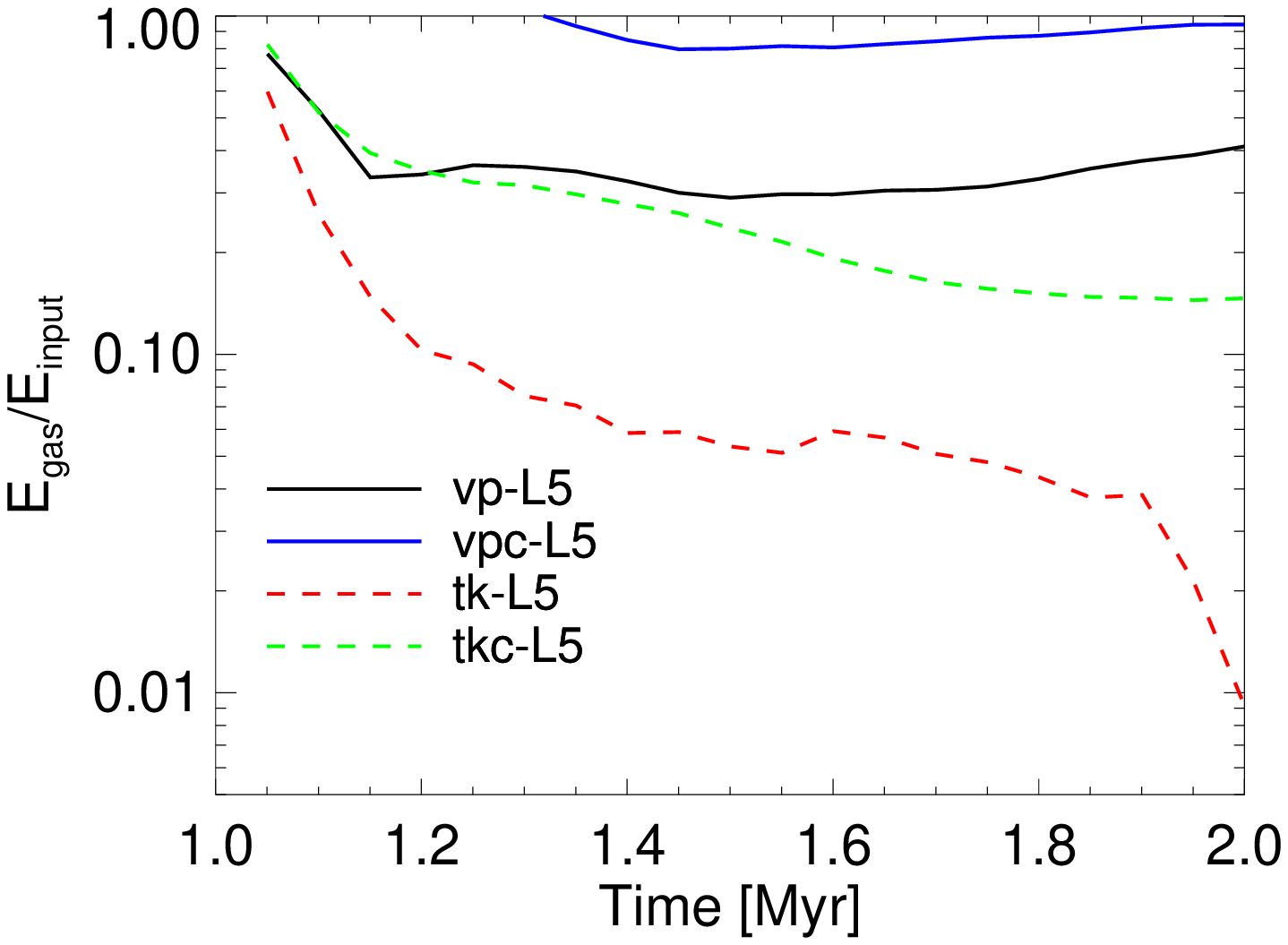}
  \caption{Time evolution of the ratio between AGN input energy and
    change in total gas energy in the four L5 simulations (top) and
    four L2 simulations (bottom). Simulations vp-L2 and vp-L5 shown in
    black solid line, vpc-L2 and vpc-L5 - blue solid, tk-L2 and tk-L5
    - red dashed, tkc-L2 and tkc-L5 - green dashed.}
  \label{fig:energy_evolution}
\end{figure}

One of the most important effects that AGN feedback has upon its host
is the energy injection, which leads to gas heating and expulsion. The
heated gas then cools down and forms structures including clumps and
filaments, as seen in the morphology plots. In order to better
understand the difference between thermal and virtual particle
feedback prescriptions, we show the phase diagrams of the four L2
simulations in Figure \ref{fig:phase_L2} and correspondingly for the
L5 simulations in Figure \ref{fig:phase_L5}. In these plots, colour
represents the density of particles in the particular parts of the
plot, with blue being lowest and red being highest. Dashed lines
indicate the bremsstrahlung cooling times.

A clear difference can be seen between the tk-L2 and vp-L2 models
(Figure \ref{fig:phase_L2}, top row). Thermal feedback creates
slightly more dense gas, which cools down quickly to the two
horizontal branches of the equilibrium temperature curve, between
$10^4$ and $10^5$~K. Meanwhile, virtual particle feedback acts upon
diffuse gas and heats it up to $10^6$-$10^8$~K. This gas has long
cooling times, $t_{\rm c} > 10^5$~yr and therefore stays hot and flows
outward. Thus the virtual particle feedback prescription allows the
gas to retain a larger fraction of the AGN input energy than the
thermal prescription. Such difference is much less pronounced in the
conical feedback runs (bottom row), leading to much more similar
morphology.

The difference between the tk-L5 and vp-L5 models is even more
pronounced (Figure \ref{fig:phase_L5}, top row), as the thermal
feedback compresses gas close to the SMBH into a very dense shell ($T
\simeq 10^4$~K, $\rho > 10^{-19}$~g~cm$^{-3}$). Meanwhile, the virtual
particle feedback creates a large population of gas with $T > 10^7$~K,
which expands rapidly outward. Once again, the conical feedback
simulations (bottom row) are much more similar to each other, with
large amounts of hot expanding gas and much less extremely dense
material.

These differences in energy retention are clearly visible when the
total gas energy is considered. In Figure \ref{fig:energy_evolution},
we show the change of total particle energy compared with input energy
as a function of time:
\begin{equation}
  f_{\rm e}\left(t\right) \equiv \frac{E_{\rm
      gas}\left(t\right)-E_{\rm gas}\left(1{\rm
      Myr}\right)}{0.05L_{\rm AGN} \left(t-1{\rm Myr}\right)};
\end{equation}  
here, the gas energy $E_{\rm gas}$ includes gravitational potential,
thermal and kinetic energy. We do not plot the results of the L1
simulations, because they are qualitatively very similar to those of
L2, shown in the top panel. The L5 models are in the bottom
panel. Virtual particle simulation results are shown in solid lines
(black for spherical, blue for conical virtual particle emission),
while the thermal feedback simulations are shown in dashed lines (red
for spherical, green for conical feedback injection).

In both cases, the conical feedback simulations lose less energy to
cooling than spherically symmetric ones. This happens because with
conical feedback, dense gas is pushed toward the midplane and the
heated gas is mostly diffuse, therefore has a low cooling rate (the
presence of hot diffuse gas is also seen in the phase plots above). In
the L2 models, this low cooling rate is still high enough to make the
gas lose most of the injected energy, leaving only a few percent as
the internal energy. However, this is still a factor few better
retention than the spherically symmetric model in the virtual particle
case; the spherically symmetric thermal feedback model has gas energy
decreasing to values even lower than those at $t=1$~Myr, as gas falls
deeper into the potential well. It is worth noting that in the tkc-L2
and vpc-L2 simulations, gas retains very similar amounts of input
energy. In the L5 simulations, the virtual particle conical feedback
model is essentially purely adiabatic, losing only $10-20\%$ of the
input energy by $2$~Myr; the thermal conical feedback model loses a
much larger fraction of its energy, since it does not create such a
large bubble. Even so, the difference between spherical and conical
feedback models is significant: energy retention improves by a factor
5 or more in the thermal model.

Even though the high-luminosity thermal feedback simulation retains
significantly less energy than its virtual particle counterpart, the
similarity in the L2 models is encouraging. Below we show that similar
trends exist in other integrated quantities.

\subsection{Inflow and outflow rates}

\begin{figure*}
  \centering
    \includegraphics[trim = 0 0 4mm 0, clip, width=0.49 \textwidth]{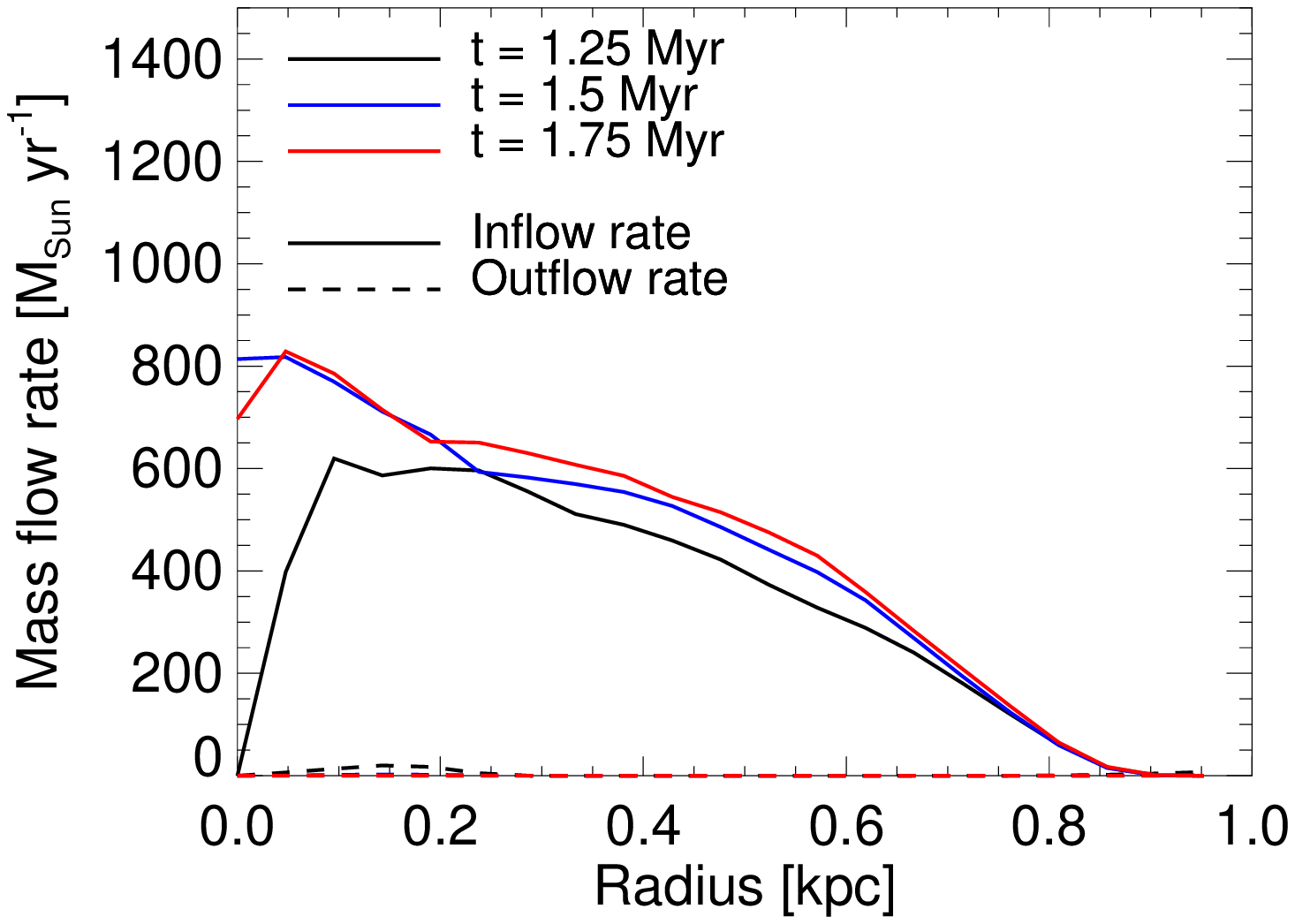}
    \includegraphics[trim = 0 0 4mm 0, clip, width=0.49 \textwidth]{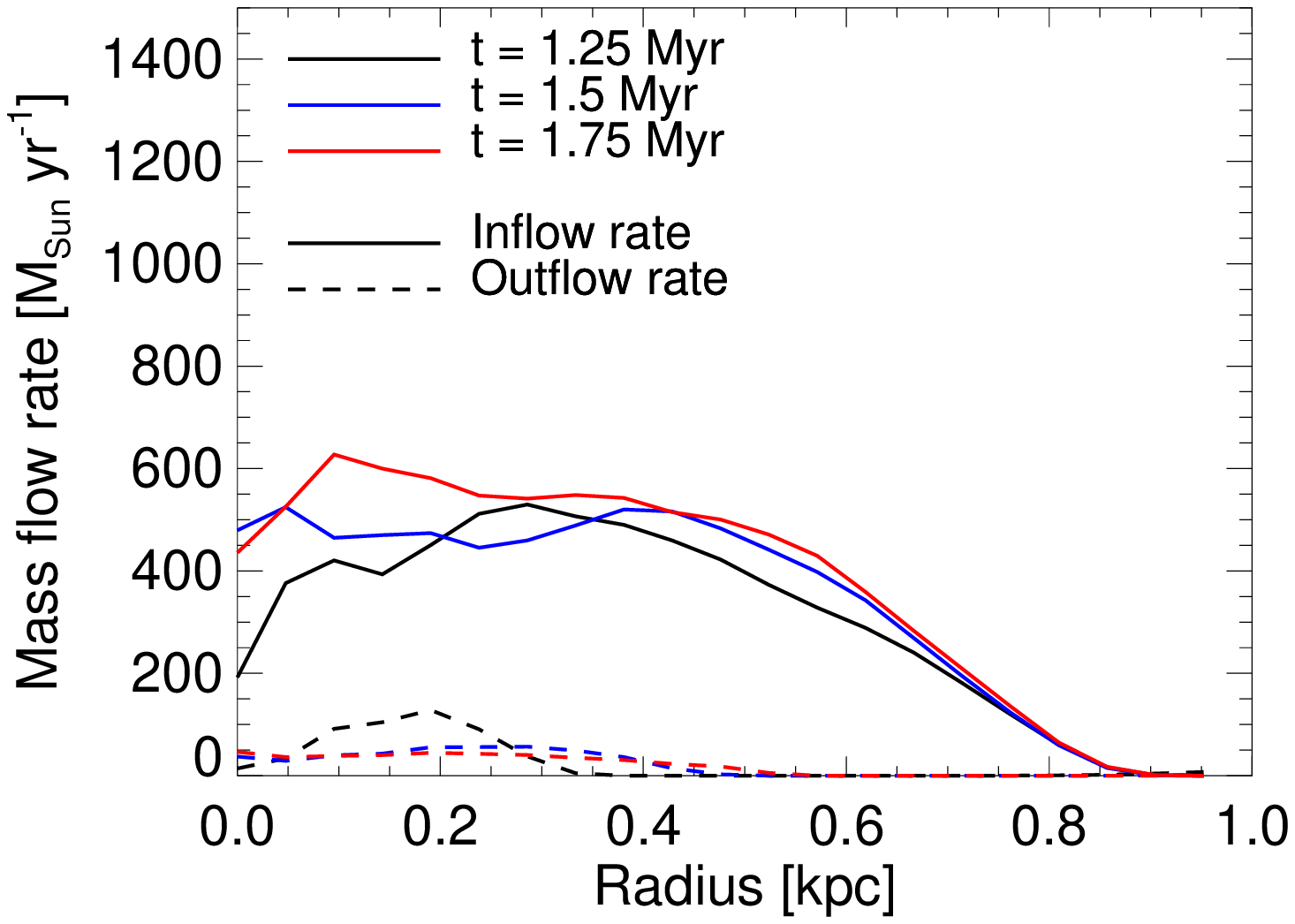}
    \includegraphics[trim = 0 0 4mm 0, clip, width=0.49 \textwidth]{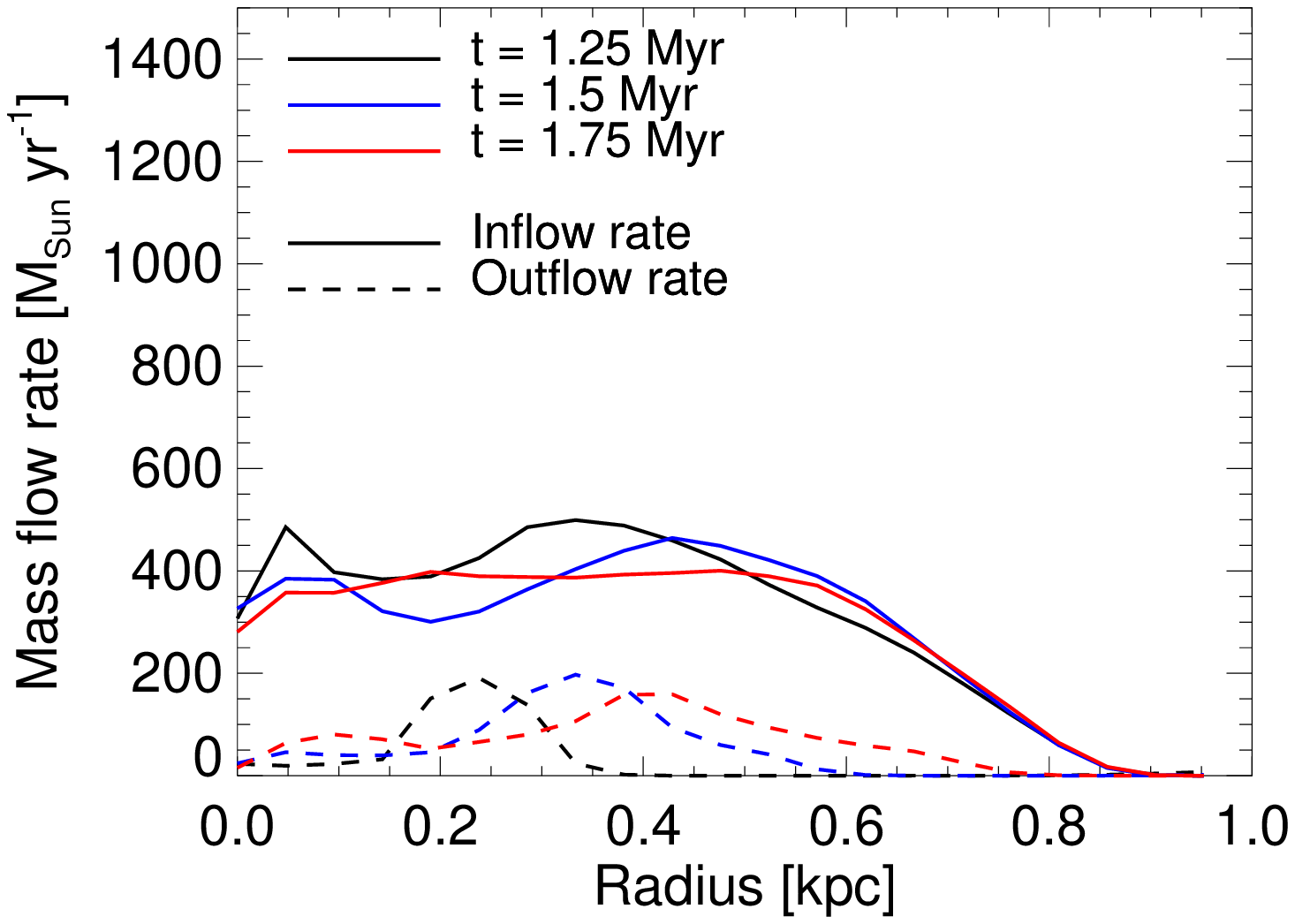}
    \includegraphics[trim = 0 0 4mm 0, clip, width=0.49 \textwidth]{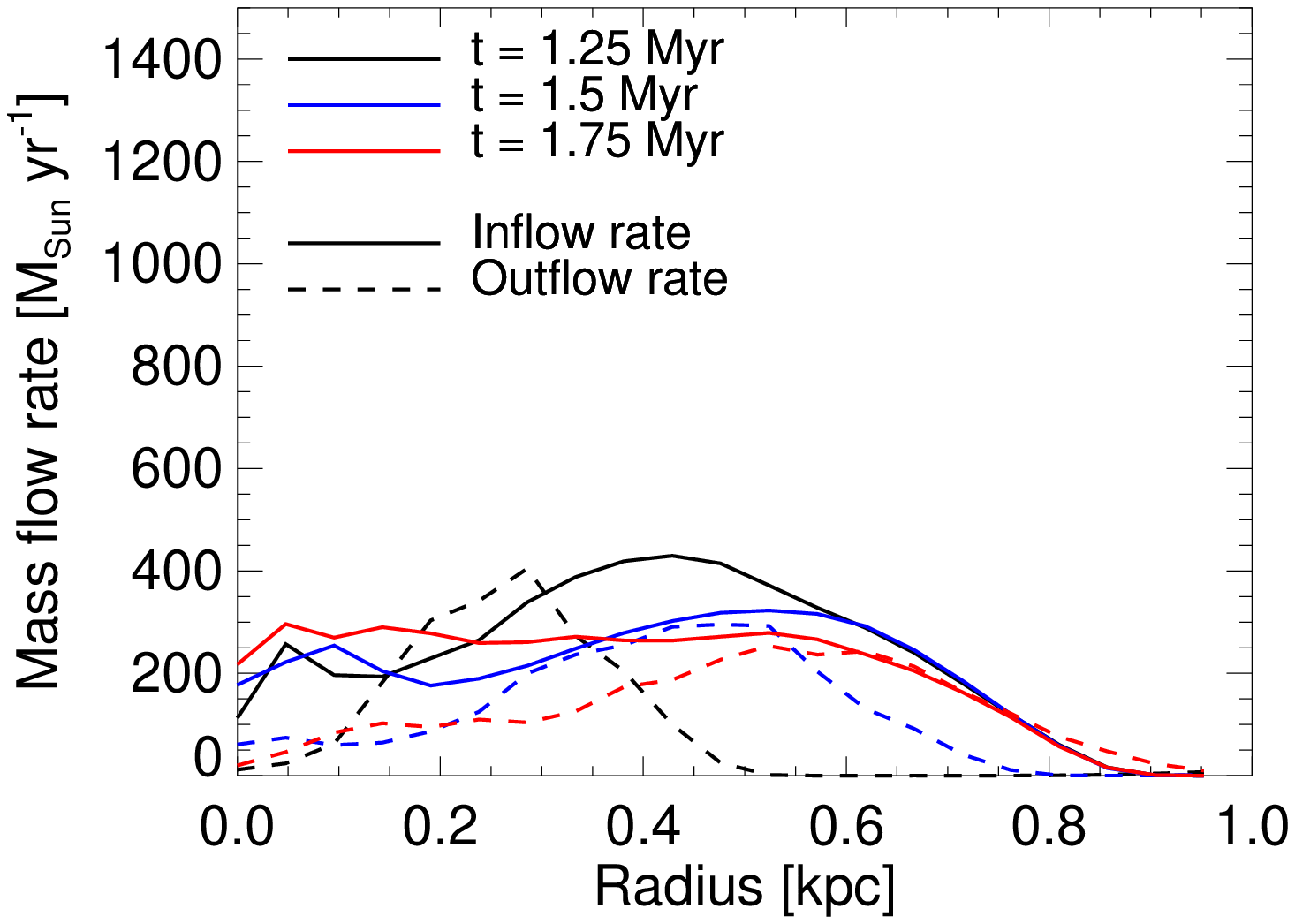}
    \caption{Radial profiles of gas outflow (dashed lines) and inflow
      (solid lines) rates for three different times in the four L2
      models: tk-L2 on the top left, vp-L2 on the top right, tkc-L2 on
      the bottom left and vpc-L2 on the bottom right.}
  \label{fig:massflow_profiles_L2}
\end{figure*}

\begin{figure*}
  \centering
    \includegraphics[trim = 0 0 4mm 0, clip, width=0.49 \textwidth]{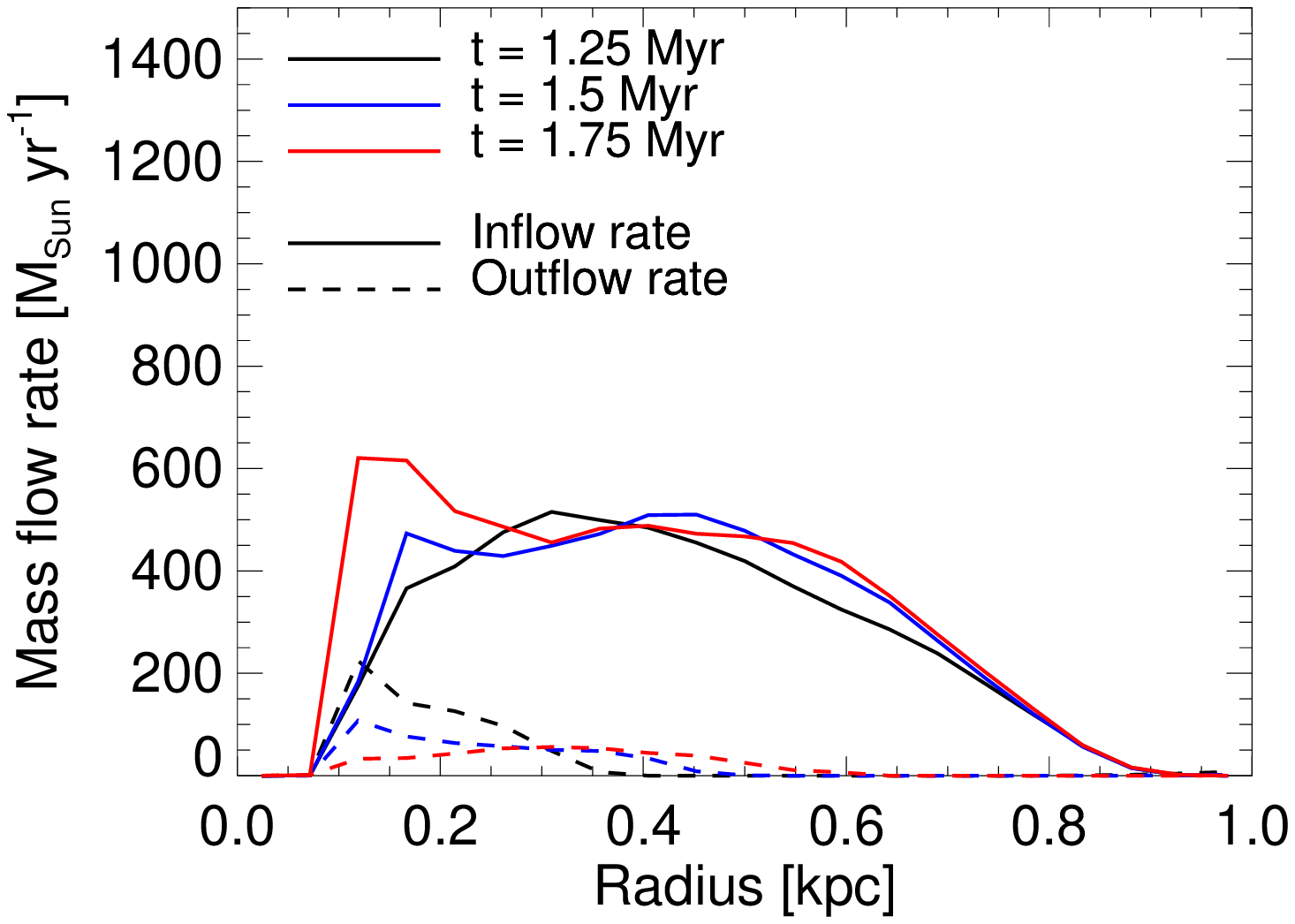}
    \includegraphics[trim = 0 0 4mm 0, clip, width=0.49 \textwidth]{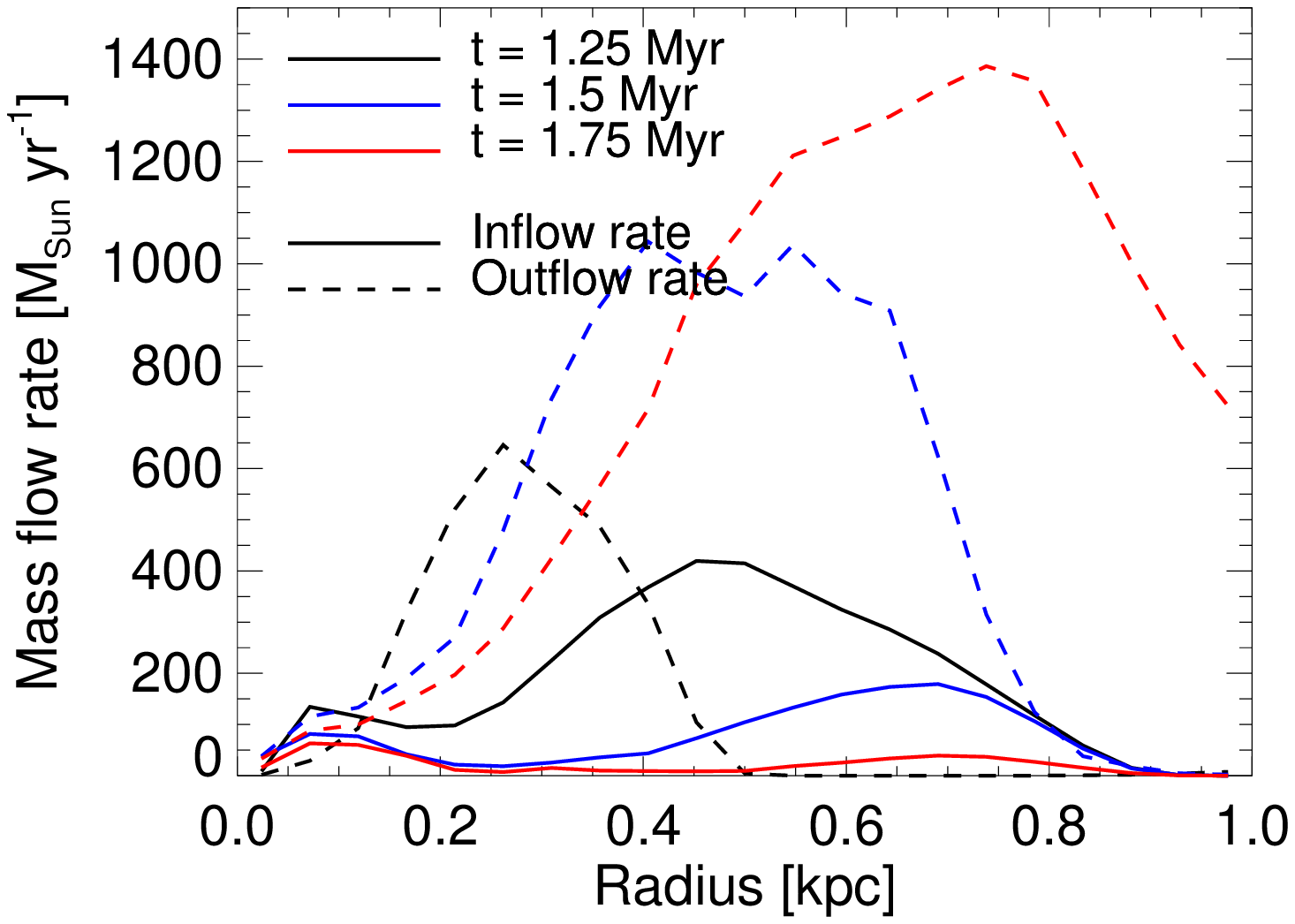}
    \includegraphics[trim = 0 0 4mm 0, clip, width=0.49 \textwidth]{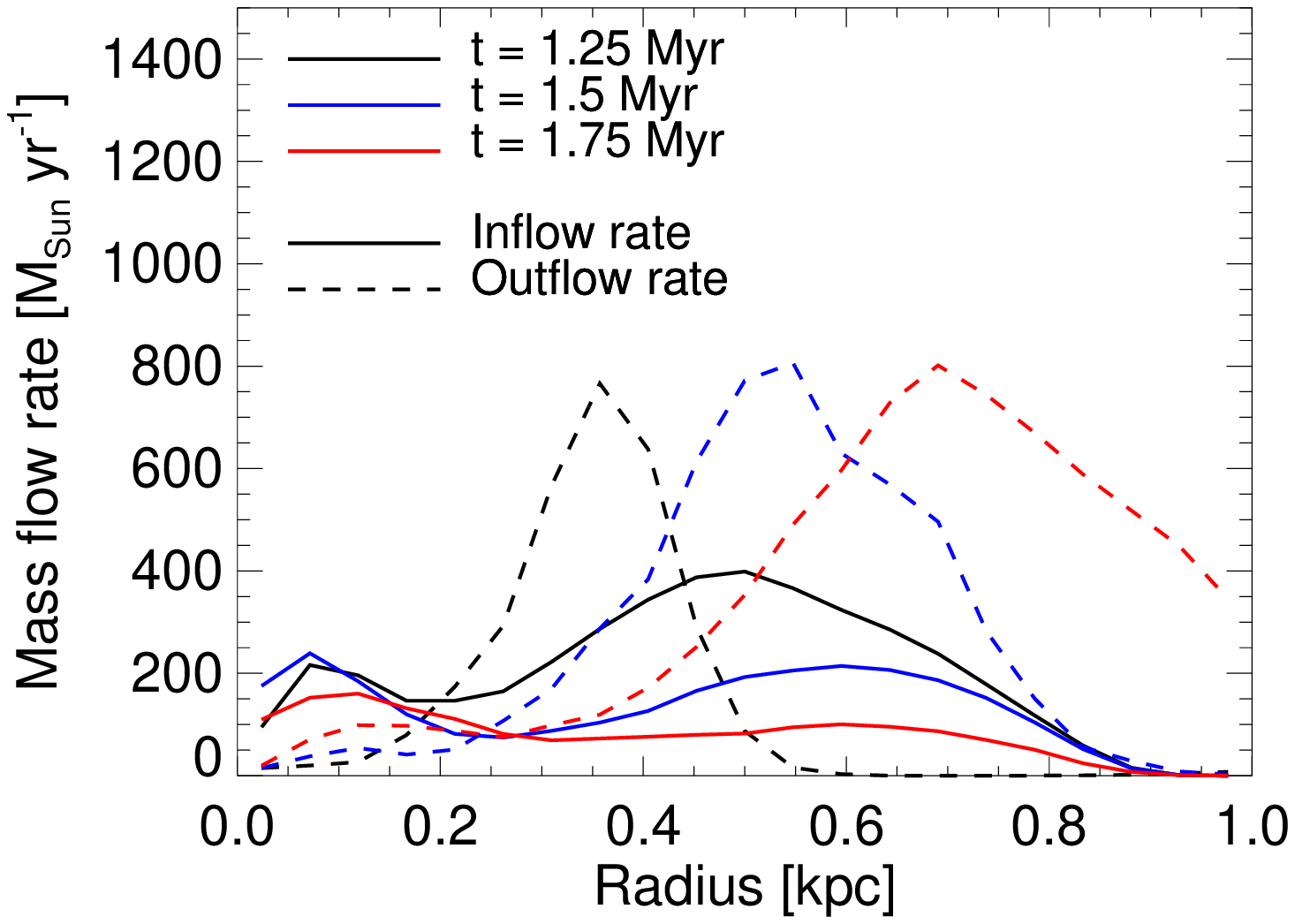}
    \includegraphics[trim = 0 0 4mm 0, clip, width=0.49 \textwidth]{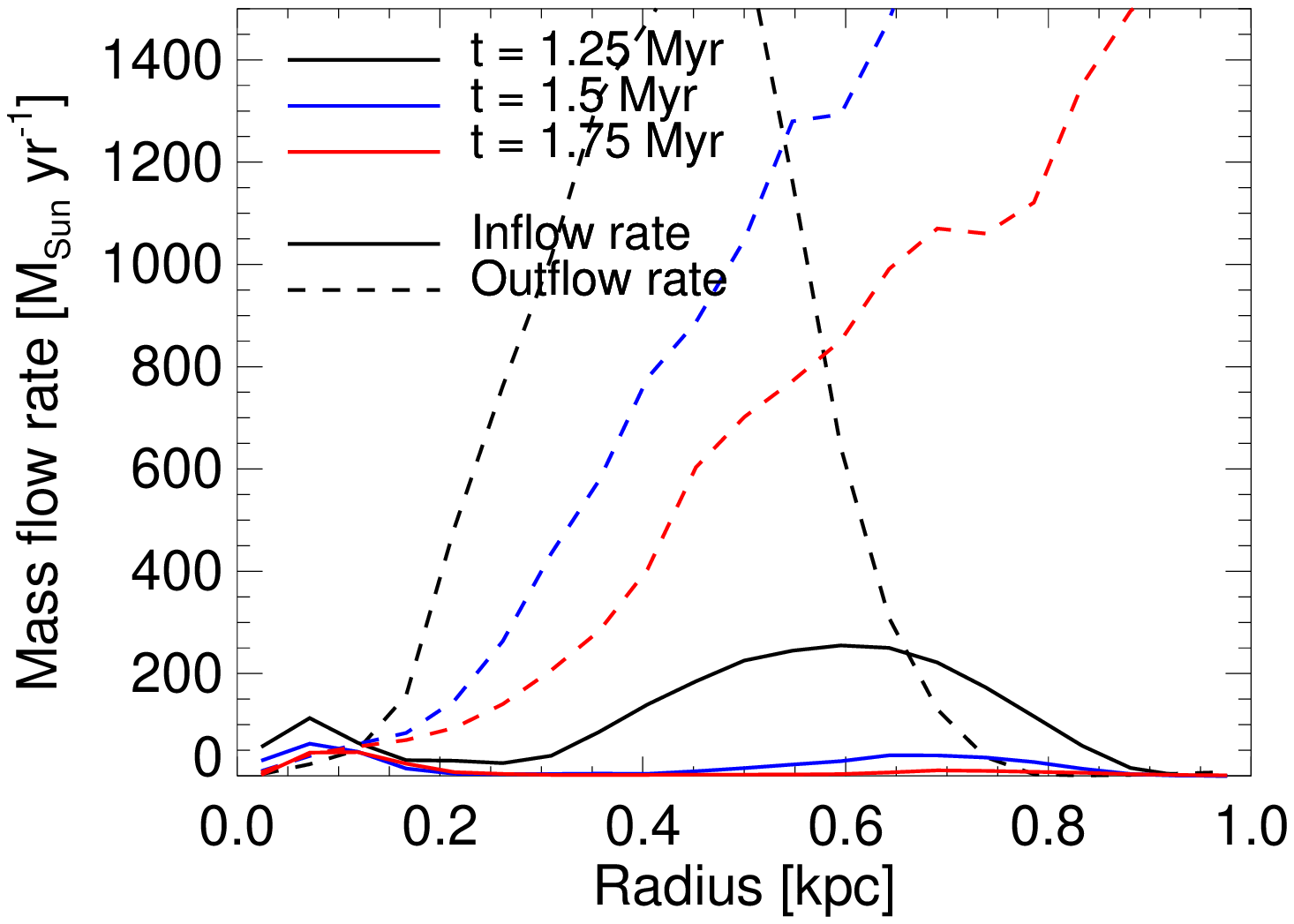}
    \caption{Same as Figure \ref{fig:massflow_profiles_L2}, but for
      the L5 simulations: tk-L5 on the top left, vp-L5 on the top
      right, tkc-L5 on the bottom left and vpc-L5 on the bottom
      right.}
  \label{fig:massflow_profiles_L5}
\end{figure*}

Another particularly important aspect of the co-evolution of SMBHs and
their galaxies is the AGN duty cycle, i.e. the frequency and duration
of AGN activity episodes. In a realistic system, one might expect AGN
feeding and feedback to occur simultaneously, at least up to some
critical AGN luminosity. Within the AGN wind feedback model, this
happens because most of the AGN wind energy is carried away by
low-density gas, allowing dense clouds to fall in toward the
SMBH. Although our simulations do not resolve the scales of SMBH
feeding, we nevertheless can investigate the ability of various
feedback prescriptions to reproduce simultaneous inflows and
outflows. For this, we plot the radial profiles of gas inflow and
outflow rates in the L2 and L5 simulations in Figures
\ref{fig:massflow_profiles_L2} and \ref{fig:massflow_profiles_L5},
respectively. The plots are made at three times for each simulation,
$0.25$~Myr, $0.5$~Myr and $0.75$~Myr after the AGN switches on. The
inflow rate is defined as
\begin{equation}
  \dot{M}_{\rm in} =
  m_{\rm SPH} \sum_{\rm i \in (v_{\rm r} < -\frac{\sigma}{2})} \frac{|v_{\rm r,i}|}{\Delta R} = 4 \pi R^2 v_{\rm r,in} \langle\rho_{\rm in}\rangle,
\end{equation}
where $\Delta R = 0.1$~kpc is the thickness of the radial bin and the
sum goes over all particles with radial velocity directed inward and
higher in magnitude than half the background velocity dispersion. The
second equality shows that the inflow rate can be expressed via the
mean inflowing gas density $\langle\rho\rangle$ and does not depend on
the choice of $\Delta R$ so long as the bin is thick enough to sample
a large number of particles. We neglect particles with small negative
velocities because those are dominated by turbulent motions rather
than inflow or outflow. The outflow rate is defined in the same way,
but the sum is made over all particles with radial velocities greater
than half $\sigma$.

The four L2 models all generally have stronger inflows than
outflows. This is expected, since the AGN luminosity is not large
enough to drive away all the gas by momentum push alone. However,
clear differences are visible among the models, with tk-L2 having
essentially zero outflow, vp-L2 having outflow rates $10-20$ times
lower than inflow, tkc-L2 having outflow rates only $2-3$ times
smaller, especially in the region $r = 0.2-0.5$~kpc, and finally
vpc-L2 having inflow and outflow rates of comparable magnitude. We see
that the qualitative difference present between the tk-L2 and vp-L2
models disappears when feedback is injected conically, even though a
notable quantitative difference remains. It is also worth noting that
the morphology of all four L2 models appears very similar in the outer
regions, since the outflows are never fast enough to extend beyond
$\sim0.5$~kpc in 1 Myr.

The L5 models display much greater differences among themselves. The
tk-L5 simulation produces very little outflow and a strong inflow,
rather similar to its lower-luminosity counterpart. However, the
inflow proceeds only to $r \simeq 0.1$~kpc and stalls there. In the
other three simulations, outflows dominate, initially in the inner
regions (black lines), but later throughout the simulation volume. The
two conical feedback simulations have qualitatively very similar
radial profiles of both inflow and outflow, echoing the similarity in
their morphologies (Figures \ref{fig:vpc_morph} and
\ref{fig:tkc_morph}). It is important to note that even though the
outflow is very powerful, the inflow rate in the central $\sim
0.15$~kpc is similar to the outflow rate, showing the the SMBH may be
fed even while producing such a massive outflow.

\begin{figure}
  \centering
    \includegraphics[trim = 0 0 4mm 0, clip, width=0.49 \textwidth]{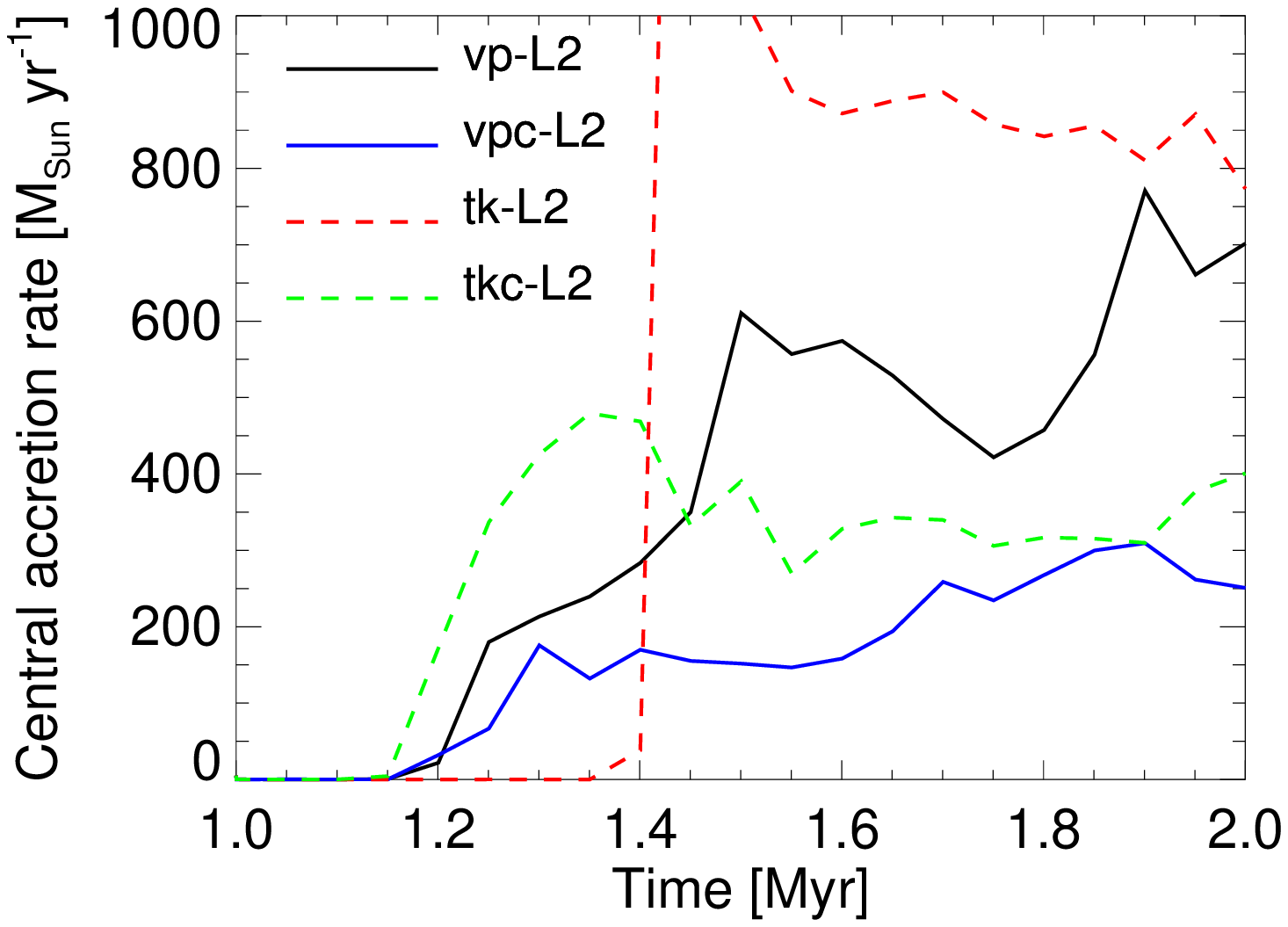}
    \includegraphics[trim = 0 0 4mm 0, clip, width=0.49 \textwidth]{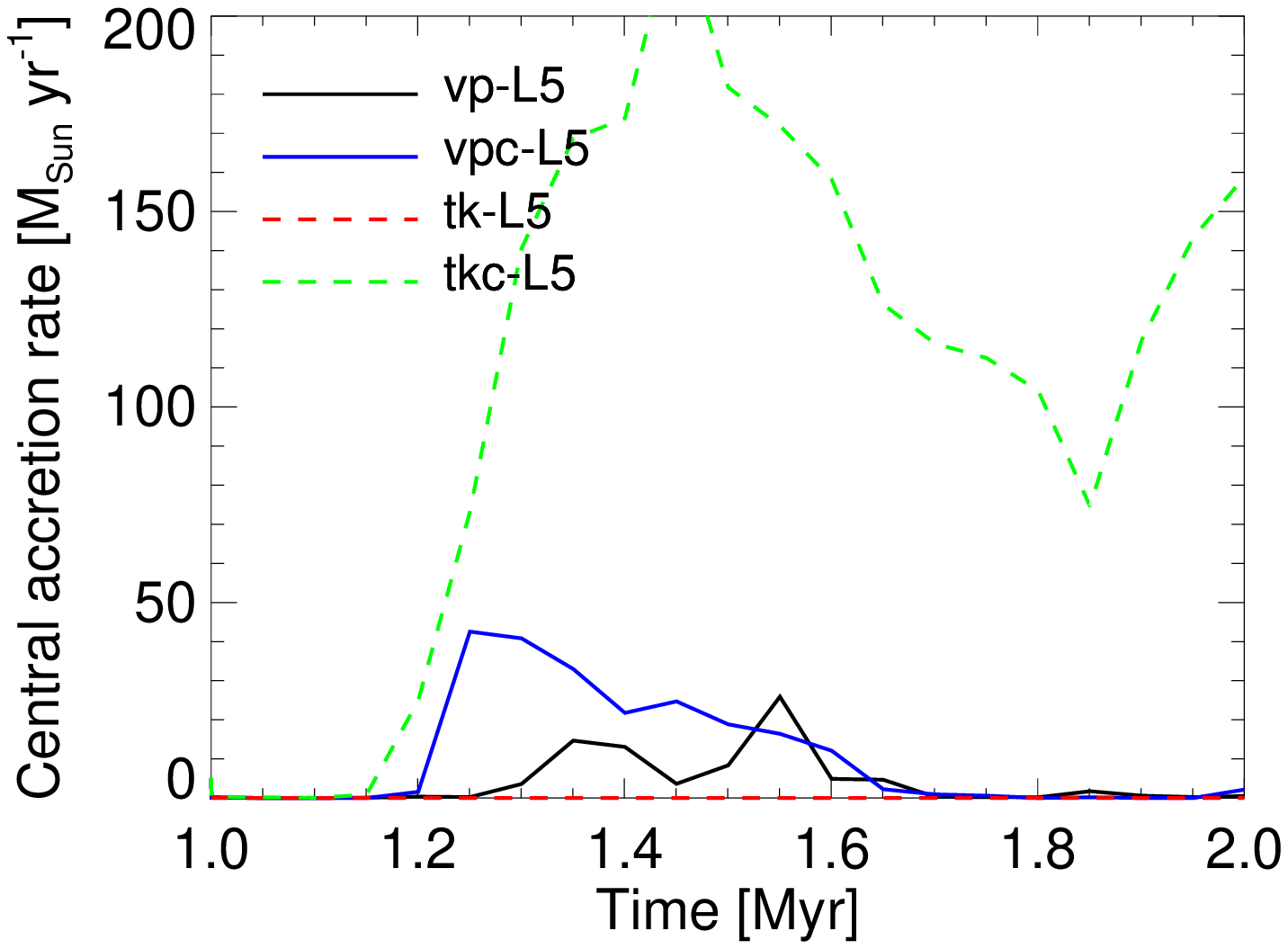}
  \caption{SMBH particle accretion rate as function of time during the
    AGN activity episode in the four L2 models (top) and L5 models
    (bottom). Colours and line styles as in Figure
    \ref{fig:energy_evolution}.}
  \label{fig:mdot_bh}
\end{figure}

\begin{figure}
  \centering
    \includegraphics[trim = 0 0 4mm 0, clip, width=0.49 \textwidth]{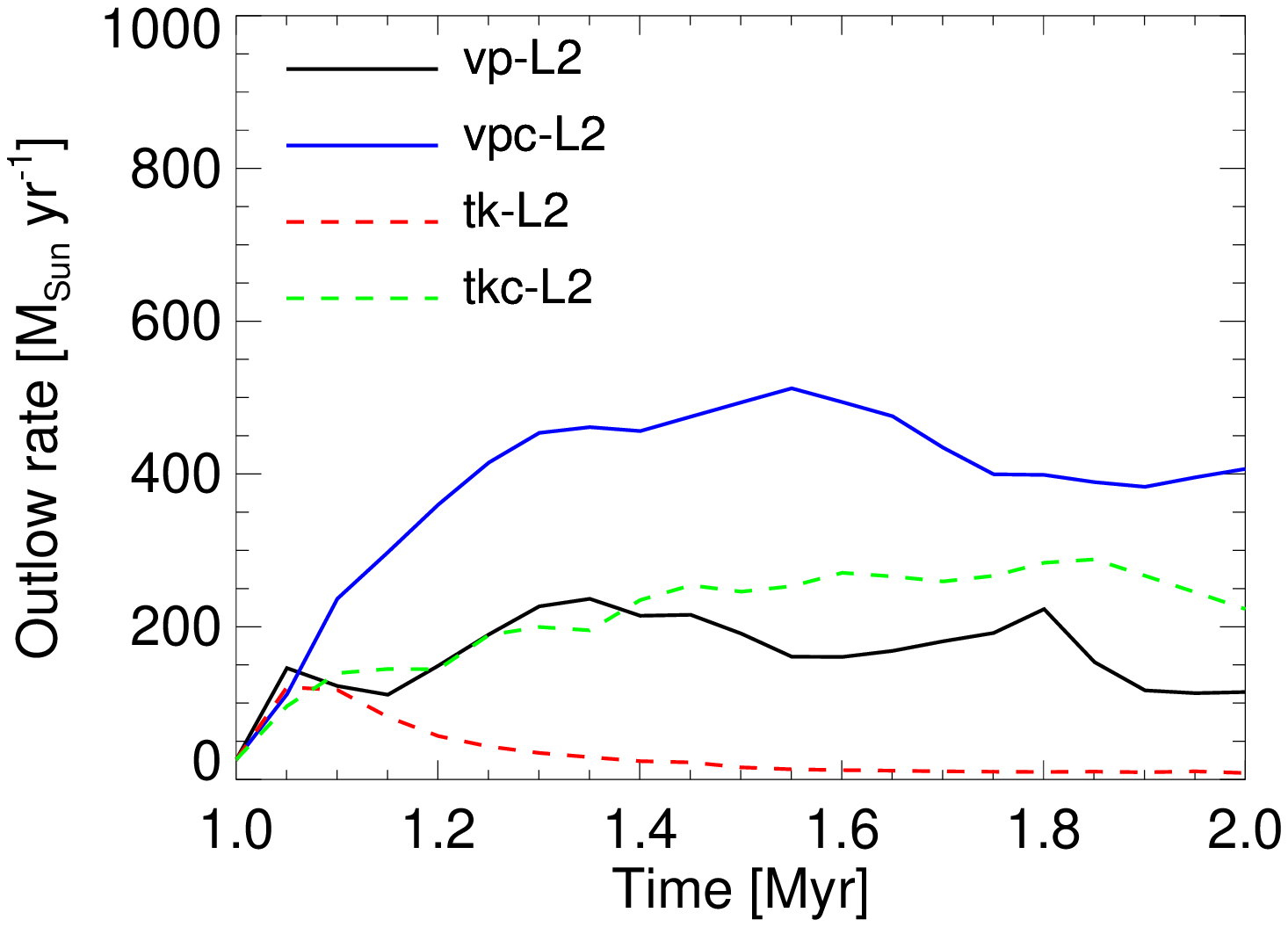}
    \includegraphics[trim = 0 0 4mm 0, clip, width=0.49 \textwidth]{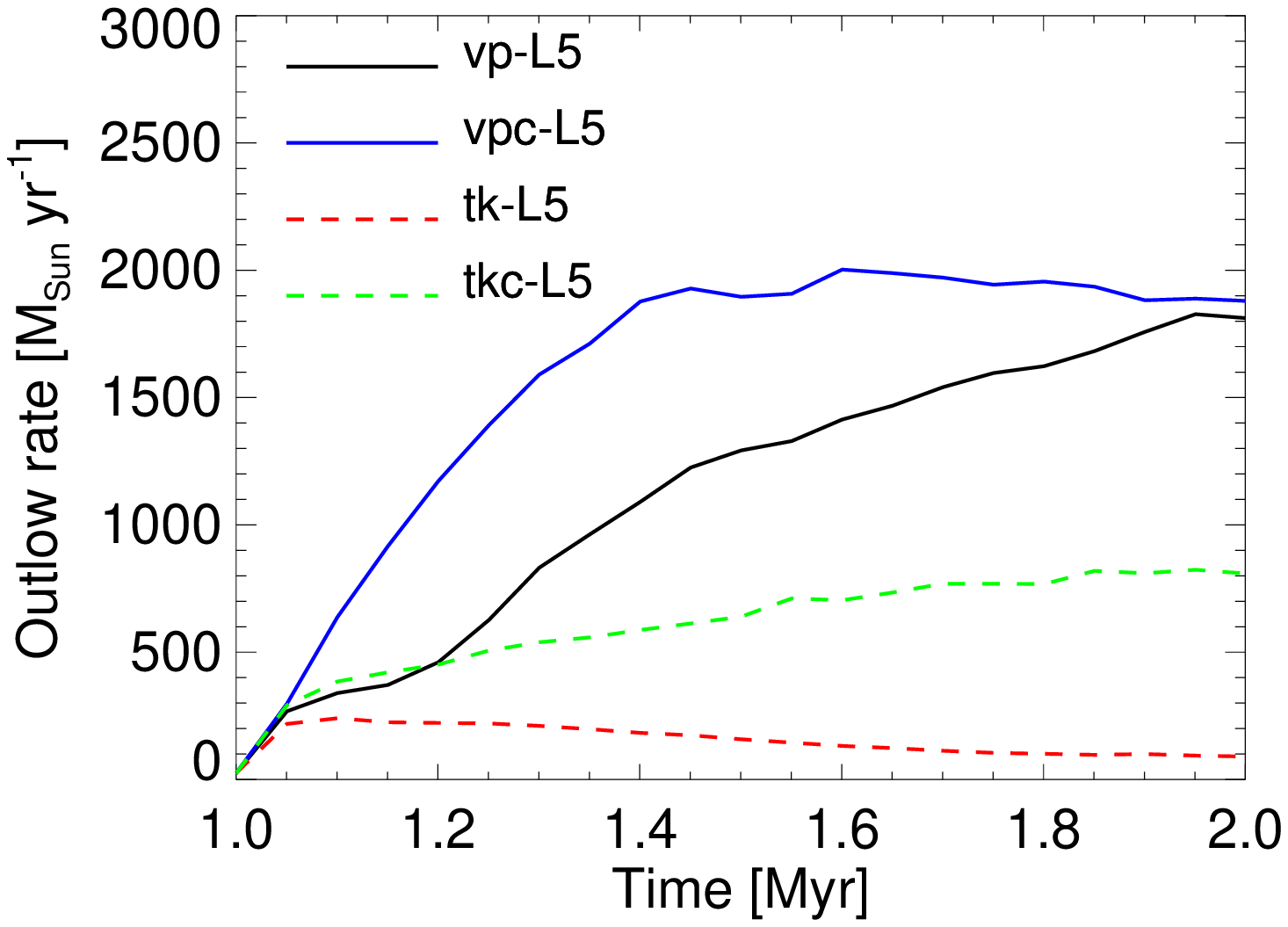}
 \caption{Gas outflow rate as function of time during the
    AGN activity episode in the four L2 models (top) and L5 models
    (bottom). Colours and line styles as in Figure
    \ref{fig:energy_evolution}.}
  \label{fig:mdot_out}
\end{figure}

We also plot the time evolution of the rate of gas particle accretion
by the SMBH particle (Figure \ref{fig:mdot_bh}) and the total rate of
gas outflow, approximated as
\begin{equation}
  \dot{M}_{\rm out} \simeq
  m_{\rm SPH} \sum_{\rm i \in (v_{\rm r} > \frac{\sigma}{2})} \frac{v_{\rm r,i}}{R_{\rm i}}
\end{equation}
(Figure \ref{fig:mdot_out}). The line colours are as in Figure
\ref{fig:energy_evolution}. In all models, there is no accretion for
the first $0.2$~Myr of AGN activity, because the gas is falling toward
the significantly reduced accretion radius of the SMBH particle.

Here, the two spherically symmetric models are more similar than when
the energy retention is considered. However, there is an important
qualitative difference. In the tk-L5 model, thermal feedback is able
to shut off accretion entirely, by keeping the gas at the edge of a
well-defined bubble (see Figure \ref{fig:tk_morph}, right panel). On
the other hand, the virtual particle model, while producing very large
outflow bubbles, also allows some gas to fall on to the SMBH
particle. The accretion rate is comparatively small, hardly exceeding
$20 \msun$~yr$^{-1}$, but the possibility of accretion is
significant. In the L2 models, the situation is reversed: thermal
feedback suppresses accretion for $\sim 0.4$~Myr, but then the outflow
bubble collapses and accretion rate rises rapidly to $>1000
\msun$~yr$^{-1}$. Meanwhile, the vp-L2 models has a steadily rising
accretion rate, without such sudden changes.

Both high-luminosity conical feedback models, unsurprisingly, allow
far more accretion on to the SMBH. Simulation vpc-L5 is more efficient
at pushing gas away than the tkc-L5 simulation, therefore the
accretion rate is significantly lower, only rising to a peak of $\sim
45 \msun$~yr$^{-1}$. The thermal feedback model, on the other hand,
allows accretion at rates approaching $200 \msun$~yr$^{-1}$ at $t =
1.25-1.6$~Myr.

In the low-luminosity models, conical feedback actually decreases the
gas inflow rate by a factor few. This happens because in these models,
outflow bubbles form and prevent accretion in some directions,
channeling accreting material toward the midplane. The tkc-L2
simulation has a higher accretion rate at first, since it is initially
more efficient in compressing the gas toward the midplane, but from $t
= 1.7$~Myr onward, the two accretion rates become very similar.

The difference among the gas outflow rates (Figure \ref{fig:mdot_out})
is far more striking. The spherical thermal feedback model is
incapable of driving a significant outflow in either low- or
high-luminosity models, as most of the AGN input energy is radiated
away, and the gas is kept in a dense, but small and hardly expanding,
shell. The spherical virtual particle simulation produces an outflow
with at least several times higher mass flow rate, and in the higher
luminosity case, the outflow rate approaches that of the conical
simulations by $t = 2$~Myr. This happens because of the complex
morphology present in the virtual particle simulation, with outflowing
gas being diffuse, and hence reaching very high velocities.

The conical feedback models produce significant outflows, essentially
because the dense gas is channeled away and diffuse gas can be
accelerated to high velocities. The mass outflow rates approach $2000
\msun$~yr$^{-1}$ in the L5 models. It is interesting that the
collimation of feedback into a cone actually increases the outflow
rate in the virtual particle simulations, most likely because conical
feedback produces more spatially distinct regions of inflow and
outflow, so there is less mixing between inflowing and outflowing gas,
which would lead to lower outflow rates. The thermal feedback model
produces a significantly weaker outflow than the virtual particle
model in the conical case as well, but the ratio between outflow rates
is much smaller than for the spherical feedback injection simulations.

\subsection{Density structure}

\begin{figure}
  \centering
    \includegraphics[trim = 0 0 4mm 0, clip, width=0.49 \textwidth]{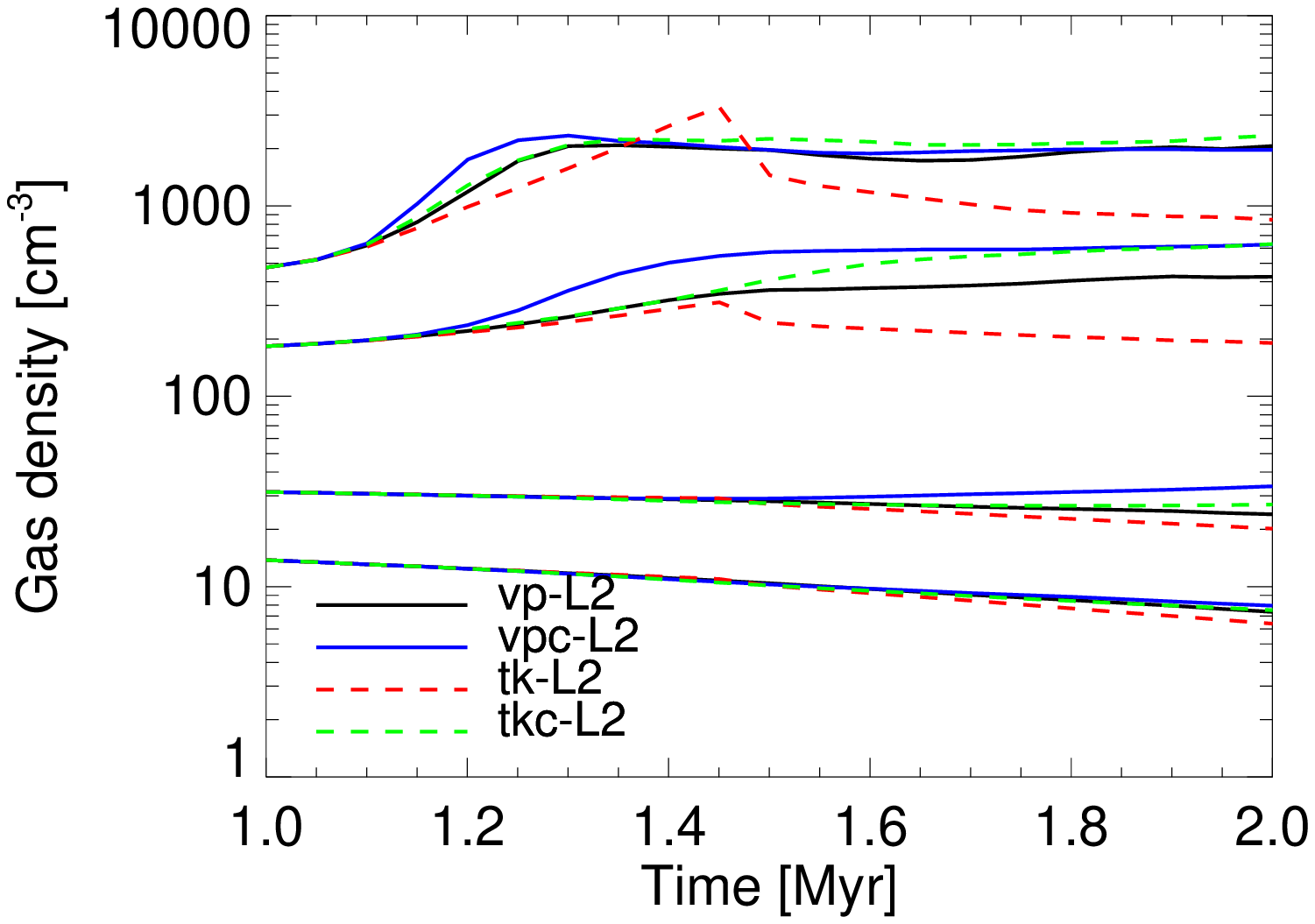}
    \includegraphics[trim = 0 0 4mm 0, clip, width=0.49 \textwidth]{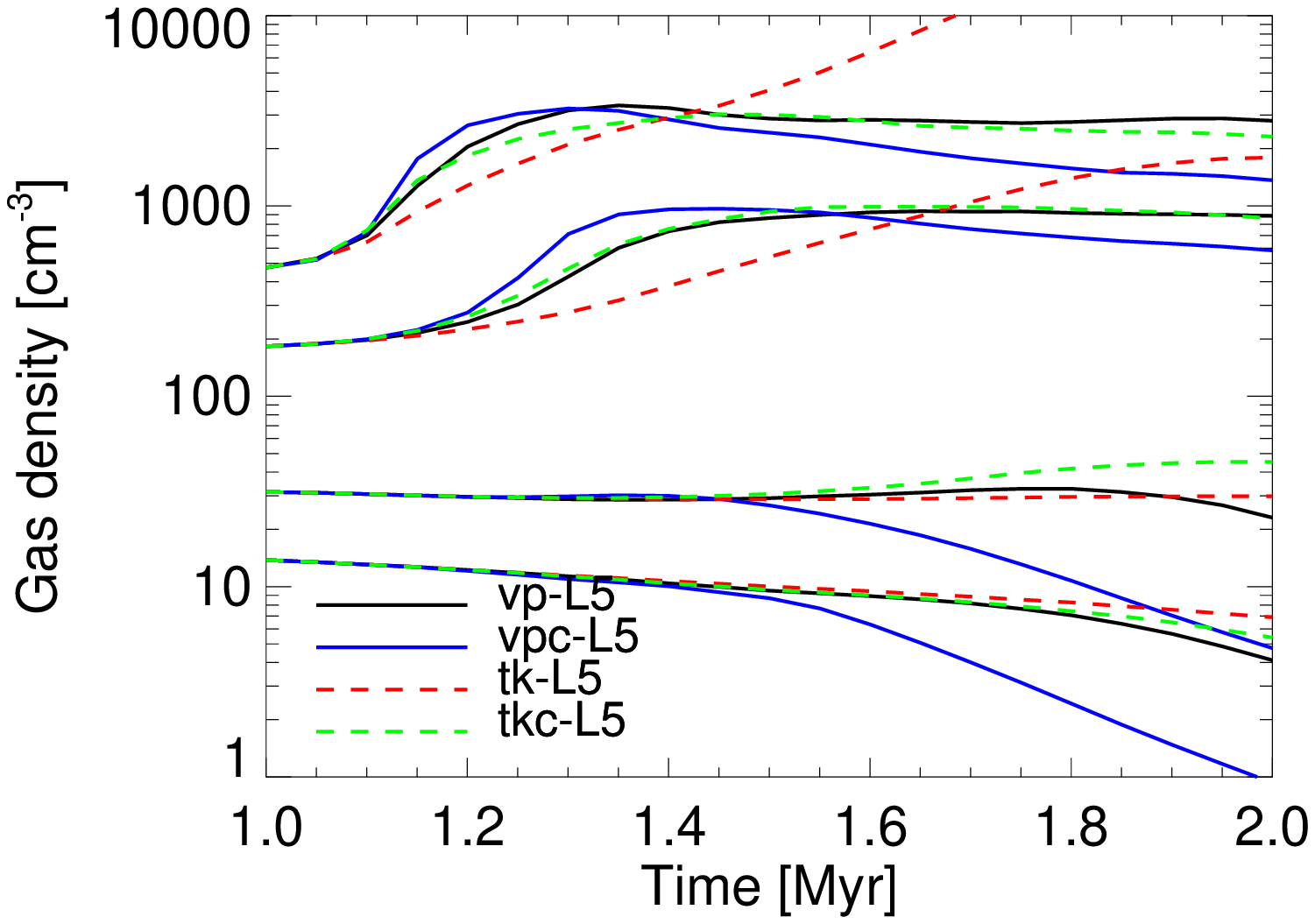}
  \caption{10th, 25th, 75th and 90th percentiles of the gas particle
    density distribution as function of time during the AGN activity
    episode in the four L2 models (top) and L5 models (bottom).
    Colours and line styles as in Figure \ref{fig:energy_evolution}.}
  \label{fig:rho_dist}
\end{figure}

In Figure \ref{fig:rho_dist}, we plot information regarding the
evolution of gas density distribution during the AGN activity
episode. More precisely, we plot the values of the 10th, 25th, 75th
and 90th percentile of the gas density distribution, in order to
evaluate how the different phases of gas (diffuse, medium and dense)
behave during the AGN activity episode. The line colours and styles
are the same as in previous time evolution figures.

We see that the diffuse gas evolves almost identically for the whole
AGN activity episode in the L2 models and for the first $\sim 0.5$~Myr
in the L5 models, mostly because most of this gas is far away from the
SMBH and is not affected by the AGN outflow at first. None of the L2
models produces a bubble extending to the outskirts of the gas sphere,
so there is little evolution of the diffuse gas in this model
throughout the time period considered; the same is true for the tk-L5
model. In the other three models, diffuse gas starts to become even
more diffuse as the outflow bubble moves past it and disperses the gas
in a larger volume.

The dense gas becomes progressively more dense, but with certain
differences among the simulations. The virtual particle models
compress the dense gas by $\sim 5-10$ times in the period between
$\sim 1.05 - 1.3$~Myr, corresponding to the formation of a complex
inflow-outflow structure with many dense clumps and filaments. The
tkc-L5 model shows a similar behaviour for the densest gas, although
the compression takes a slightly longer time. Once this structure is
in place, however, density does not increase further, as the densest
material either turns into sink particles or is accreted on to the
SMBH. The tk-L5 model, on the other hand, shows essentially
unrestricted density growth, because here, the densest gas is also the
warmest, because it receives all the AGN energy input, and therefore
cannot fragment. The tk-L2 model behaves in the same fashion
initially, but later the dense gas collapses on to the SMBH and is
accreted, resulting in a drop in the densest gas density values.

\section{Discussion}\label{sec:discuss}

\subsection{Summary of results}

Our suite of 13 simulations allowed us to explore the effect that the
subgrid AGN feedback prescription has upon the evolution of a
turbulent gas sphere. Our main results are as follows:

\begin{itemize}

  \item Spherically symmetric thermal AGN feedback is unable to
    reproduce the complex structures found in virtual particle
    models. The reason is that the thermal feedback prescription
    injects energy only into the surrounding gas, creating a
    well-defined bubble around the SMBH.

  \item The virtual particle prescription allows for simultaneous gas
    inflow and outflow, as well as formation of dense gas filaments
    and clumps, which are not found in the thermal feedback case.

  \item Replacing both prescriptions with biconical versions results
    in a marked improvement in their agreement. In particular, the
    morphologies of the two models then become very similar, with
    clear outflow bubbles coexisting with inflow along the
    equatorial plane.

  \item The conical feedback models are qualitatively similar to
    the corresponding spherical virtual particle simulations, but
    quantitatively very different from both spherical models: the
    outflow bubbles produced by conical feedback are generally larger,
    the outflow rates higher and much more of the AGN input energy is
    retained by the gas.

  \item The fraction of AGN input energy retained in the gas is very
    low in the spherically symmetric thermal feedback simulation, even
    at high AGN luminosity, but increases to $\sim 20-30\%$ in the
    conical high-luminosity one. In the virtual particle models, the
    energy retention fraction increases to an ever higher value of
    $70-80\%$.

  \item Spherically symmetric thermal feedback either does not allow
    any accretion on to the SMBH, or allows most of the gas to fall
    in. It also produces only a weak outflow. Virtual particle and
    conical models produce both significant inflow and outflow at the
    same time.

\end{itemize}

Although the set up of our simulations is strongly idealised, the
results have several implications for the understanding of AGN
activity and its effects upon the host galaxy. We discuss those
implications below.

\subsection{Major issues with thermal feedback models}

As mentioned in the Introduction and Model sections, the commonly-used
thermal-kinetic AGN feedback models suffer from a number of
drawbacks. The most important drawback is the assumed spherical
symmetry, i.e. energy injection into gas close to the SMBH. If it
happens that there is gas only to one side of the SMBH, all the energy
will be injected into this gas, rather than spread around
symmetrically. This creates a paradoxical situation where assumed
spherical symmetry results in energy injection which is often very far
from symmetric.

The problem with this energy injection in generating outflows lies in
the fact that once any parcel of gas escapes far from the SMBH, it is
no longer affected by feedback. Therefore, as far as most gas
particles in the simulation are concerned, feedback is very
intermittent, and this leads to much weaker outflows than would be
possible if feedback acted upon a gas parcel for a prolonged period of
time. A small amount of gas does experience this prolonged feedback,
however, it gets squashed between AGN feedback on one side and
inflowing gas on the other. Gas density therefore increases, cooling
becomes more efficient and progressively more of the feedback energy
is lost to radiation rather than used to drive the outflow. Therefore,
the resulting AGN outflow is much weaker than it would otherwise and
more energy is required to remove the gas from the host galaxy than it
would be if these multi-phase effects were treated properly.

These issues do not necessarily impact the ability of cosmological
simulations to reproduce the mass functions and growth histories of
present-day galaxies and SMBHs, even though the latter requires
appropriate tuning of feedback parameters \citep{Schaye2015MNRAS,
  Crain2015MNRAS}. AGN accretion and feedback self-regulate in these
simulations, so that a less efficient AGN feedback allows more
accretion which in turn increases the outflow rate, having a
negligible effect upon the long-term evolution of the total stellar
mass in the galaxy \citep{Sijacki2007MNRAS, Schaye2010MNRAS,
  Schaye2015MNRAS}. However, as such models, as well as models of
isolated galaxies, become more detailed, they start reproducing other,
smaller-scale, processes and structures, such as the phase structure
of the gas, the spatially-resolved history of star formation, the
duration of AGN episodes, and so on. All of these, and many other,
properties of the galaxy are affected by the details of how AGN
feedback interacts with the ISM, and therefore understanding the
limitations imposed by the numerical feedback scheme is of paramount
importance when interpreting the results of these models.

Ideally, one would like to minimize the limitations of the numerical
method. Our proposed scheme of injecting feedback energy into a cone
alleviates the two biggest issues of thermal feedback prescriptions
described above. While it does not really resolve the interaction of
the AGN outflow with the multiphase gas, it creates a possibility for
the gas to remain in two phases (cold dense inflowing and hot diffuse
outflowing) while interacting with the outflow. Dense gas is quickly
channeled to the bottom of the effective potential well, i.e. to the
regions where feedback doesn't counteract gravity. Once there, dense
gas can form filaments which feed the SMBH or fragment into
self-gravitating clumps. Meanwhile, diffuse gas gets heated and pushed
by the feedback energy injection and forms a massive outflow which
cools slowly due to low gas density. This results in an adiabatic
outflow, consistent with the analytical predictions of the wind
feedback model.

\subsection{Possible improvements to cosmological simulations}

Given that a conical thermal feedback prescription is capable of
resolving a lot of the complexity of simultaneous inflows and outflows
with little extra computational cost, we suggest that it may be
interesting to explore its effects in cosmological simulations, which
self-consistently track the growth and energy release from AGN over
long timescales. This would allow one to constrain the AGN growth
scenarios, the $M-\sigma$ relation, the effects of AGN outflows upon
galaxies and clusters, the triggering of star formation, as well as
the history of SMBH growth, with better precision than is currently
available \citep{Schaye2015MNRAS}. One particularly interesting
consequence might be that the presence of simultaneous inflows and
outflows around SMBH change the qualitative behaviour of AGN
self-regulation (see Section \ref{sec:selfreg} below).

A few important aspects have to be considered when implementing this
scheme into cosmological simulations. The first is numerical
resolution, which is typically much worse in cosmological simulations
than it is in the simulations presented here. In a previous paper
\citep{Bourne2015MNRAS}, we investigated the importance of resolution
upon the efficiency of feedback and found that low-resolution
simulations produce feedback that is much more negative. This happens
in part because those simulations are unable to resolve the complex
structures developing in the turbulent ISM. With a conical feedback
prescription, it should be easier to resolve those structures, since
the cavities and filaments are larger (compare the right panels in
Figures \ref{fig:vp_morph} and \ref{fig:vpc_morph}). However, we may
still expect a weaker overall effect of AGN feedback due to the fact
that the gas density contrast is not resolved as well, leading to
overcooling of the hot gas in the bubbles.

Another aspect is defining the appropriate direction of the cone's
axis. We suggest that this axis should coincide with the axis of
rotation of the SMBH particle. Rotation can be tracked by tracking the
angular momentum magnitude and direction of the infalling gas, and the
plane of the accretion disc should be perpendicular to this axis of
rotation. Since the AGN outflow originates in the disc, it is
reasonable to assume that it goes in the polar direction
\citep{Feldmeier1999ApJ}. The opening angle of the wind cone is a free
parameter, which might be constrained by more detailed models, or
investigated empirically. However, such an investigation is beyond the
scope of this paper.

Finally, cosmological simulations contain galaxies which have varied
structures, including discs, large clumps and cavities created by
previous outflows. The interaction of a new outflow with these
structures may be much more complex than in our simulations presented
above. For example, one directed conical outflow might intercept a
dense clump and lose a large fraction of its energy, while another may
expand into a pre-existing cavity, having little effect upon the mass
budget in its host galaxy. A more thorough investigation of feedback
prescriptions in a more realistic galaxy model should be done before
implementing them into a cosmological simulation. We intend to perform
such tests in a future publication.

\subsection{Self-regulation of AGN activity} \label{sec:selfreg}

Our simulations use a fixed AGN luminosity rather than keeping it tied
to the central particle accretion rate. Given that the central
accretion rates in different simulations can differ by orders of
magnitude (see Figure \ref{fig:mdot_bh}), this may seem as a
significant drawback of the numerical setup. However, we feel that
simulations addressing just the feedback aspect of AGN activity are
still very useful since the models with feedback determined by BH
accretion rate employ a number of assumptions about this rate. Given
the non-linear accretion-feedback connection, it would be hard to
arrive at definite conclusions about the feedback efficiency in these
more complex and less constrained simulations. In addition, the actual
expected evolution of the system is somewhat more complicated than the
accretion rate of the central particle would suggest. Infalling gas
contains some angular momentum, and therefore forms an accretion disc
around the SMBH, rather than falling in directly. Given that in the
simulations the accretion rate is measured at a radius of 10 pc, most
of this accreting gas will take far longer than $1$~Myr to reach the
SMBH, by which time the activity episode in our test run is
over. Additionally, some of this gas will fragment and form a nuclear
star cluster rather than feeding the SMBH \citep{Nayakshin2006MNRASb},
further reducing the impact upon the AGN luminosity. Therefore, our
simulations may be seen as an investigation of how a large-scale gas
distribution is affected by AGN activity caused by accreting a cloud
of predetermined mass, from $M_{\rm accr} = 2.2\times 10^6 \msun$ (L1
simulations) to $M_{\rm accr} = 1.1\times 10^7 \msun$ (L5
simulations).

In a more realistic simulation, long-term self-regulation of AGN
activity becomes important. Here, the temporal evolution of
simulations using the thermal-kinetic and virtual particle feedback
prescriptions might be very different from one another. When the AGN
luminosity increases, simulations using the spherically-symmetric
thermal-kinetic AGN feedback prescription abruptly change from a
situation where no outflow is occurring to one where the outflow
completely shuts off AGN accretion. This latter situation may reverse
after some time, as the material piles up at the edge of the outflow
bubble, and its weight may overcome the force produced by AGN
feedback, leading to collapse and resuming of accretion. However,
while the bubble holds, there is no accretion. Conversely, while there
is accretion, no outflow bubble can exist. This ``all-or-nothing''
situation places tight constraints on the regulation of AGN accretion
rate. When the AGN is fed at a large enough rate, an outflow forms and
shuts off accretion. As the accretion rate drops to zero (with perhaps
some delay imposed in the simulation to mimic the draining of the
sub-resolution accretion disc), AGN luminosity also decreases and
feedback stops, allowing accretion to resume. We may therefore expect
AGN activity to keep switching on and off, with the duration of the
``on'' phase comparable to the viscous timescale of the accretion disc
($t_{\rm on} \sim 10^5 - 10^6$~yr) and the ``off'' phase lasting for
as long as it takes for gas to fall back to the SMBH ($t_{\rm off} >
10^6$~yr). These timescales are comparable to some estimates of the
lifetimes of AGN phase \citep{Schawinski2015arXiv}, but this is only
coincidental, since the processes described here happen due to
numerical, rather than physical, reasons.

The virtual particle prescription, as well as both conical
prescriptions explored in this paper, allow simultaneous inflow and
outflow to coexist. At some point, the larger gas reservoir gets
depleted and accretion switches off, leading to a longer period of
inactivity than in the case of spherical thermal feedback. This
simulation therefore does not manifest an artificial short-timescale
``flickering'' and can be used to investigate the physical reasons of
the switching on and off of AGN activity.

On the other hand, some integrated parameters, such as the total mass
accreted by the SMBH or expelled from the galaxy, may not differ very
much between thermal-kinetic and virtual particle feedback
simulations. The reason for this is simply that on long timescales
SMBH accretion tends to self-regulate so that the total energy release
into the surrounding medium is just enough to balance gravity
\citep{Booth2009MNRAS}. A thermal-kinetic feedback prescription would
lead to faster accretion followed by a stronger burst of AGN activity,
which leads to a larger feedback bubble and shutting off of accretion
for a longer time. The virtual particle prescription would result in a
more uniform accretion and outflow rate, with neither undergoing such
strong changes as in the thermal-kinetic feedback model. As a result,
eventually the results of the two simulations might look more similar
to each other than might seem from our simulation results. Still, on
short timescales, the morphology and temporal variability of inflow
and outflow rates would differ significantly among the simulations.

The possibility of having prolonged periods of AGN activity coincident
with SMBH growth via accretion also has an implication for the growth
of the very first SMBHs. These black holes are known to have masses
exceeding $10^9 \msun$ by $z = 6$, when the Universe was $<10^9$~yr
old \citep[e.g.][]{Wu2015Natur}. If these SMBHs grew from stellar-mass
progenitors, their growth rate must have been close to the Eddington
limit for most of their lives \citep{King2006MNRAS}. The possibility
of having rapid gas inflows during the AGN phase suggests that this
might be easier than previously thought and eliminates one of the
potential drawbacks of this growth mechanism in explaining the large
observed masses. We note, however, that it is possible to grow black
holes to the required masses at high-z even with the typical
thermal-kinetic AGN feedback prescription \citep{DiMatteo2012ApJ}, at
least for seed SMBH masses of $10^5 \msun$. Therefore a change in our
understanding of accretion and feedback is not necessary to explain
the high-z quasars.

\subsection{Implications for positive AGN feedback}

Thermal feedback simulations, both spherically symmetric and conical,
eliminate some of the structure present in turbulent gas. Meanwhile,
virtual particle simulations, and to some extent the conical thermal
feedback simulations, enhance some of the structures, by enveloping
dense gas in hot diffuse bubbles and compressing it more than
self-gravity alone would. This leads to formation of more clumps and
more efficient star formation within them \citep{Zubovas2014MNRASc},
resulting in a higher rate of fragmentation and star formation. In
other words, AGN activity has a positive feedback upon star formation
in the host galaxy.

This effect has been explored in previous numerical works.
\citet{Nayakshin2012MNRAS} showed that the outflowing material can
fragment as it cools down, leading to potentially rapid star formation
and stars ejected from host galaxies. \citet{Gaibler2012MNRAS}
explored the interaction of an AGN jet with a clumpy ISM and found
that the jet compresses dense gas, prevents adiabatic re-expansion of
the clumps, and enhances the SFR of the whole galaxy. There is also
tentative observational support that galaxies with more luminous AGN
have higher star formation rates and efficiencies
\citep[e.g.,][]{Wei2010ApJ, Wang2010ApJb}, although the causal
connection is not certain.

We chose not to consider in detail the fragmentation of gas in our
simulations, as that is not the main point of the present
study. Nevertheless, we find qualitatively similar effects, but,
importantly, they are much more significant in the virtual particle
feedback models, where the multiphase ISM is affected by feedback
based on its density. In a more realistic simulation, the effects of
AGN outflow, and of the numerical prescription used to track its
interaction with the ISM, might significantly affect the fragmentation
rate and spatial locations in the galaxy. We suggest that using a more
detailed AGN feedback prescription, such as the conical thermal
feedback one, would allow investigating this positive feedback in a
far more realistic way.

\subsection{How realistic is the virtual particle simulation?}

One implicit assumption we made so far when analysing the results is
that the virtual particle feedback prescription is a good
representation of reality. The method certainly has several advantages
over the thermal feedback model. It is truly isotropic, i.e. feedback
energy is emitted in all directions from the AGN and interacts with
the ISM in all directions independently of the shape of the cavity
surrounding the AGN. In addition, the model, by construction, accounts
for the different optical depths of gas in different directions,
leading to situations where a dense clump can shield the gas behind
itself from being blown away.

On the other hand, the model suffers from some shortcomings as
well. The interaction cross-section between virtual particles and SPH
particles is a free parameter of the model. This means that the depth
to which virtual particles penetrate the gas can vary significantly
between simulations. As a result, dense gas clumps can sometimes be
obliterated unphysically rapidly. Conversely, in other cases, gas
might be compressed too strongly where a gentler push would be more
realistic and represent the many filaments forming along the contact
discontinuity between the AGN wind and the shocked ISM.

Another drawback is that the virtual particles are assumed to always
propagate radially out from the AGN. This is unrealistic in cases of
uneven density, where a steep density gradient may be present in a
direction not parallel to the direction of propagation of the virtual
particle. In this case, a realistic interaction would have the virtual
particle flying off at an angle to its initial direction, something
which the present model does not allow. This drawback may have many
subtle effects in how the gas morphology is affected by AGN activity.

Given these drawbacks, we stress that the results of the virtual
particle simulations should not be considered as perfect
representations of reality. However, given that they reproduce a far
more complex morphology of the turbulent ISM and properties of
outflows in agreement with observations, we believe that their results
represent real galaxies better than those of the spherical thermal
feedback models.

\section{Conclusion}\label{sec:concl}

In this paper, we presented a number of simulations of AGN feedback
affecting a turbulent gas sphere. The simulations are designed to show
the impact of different subgrid recipes of AGN feedback upon the
evolution of the whole system. We show that the commonly-used
spherically-symmetric thermal AGN feedback prescription is unable to
reproduce the variety of complex structures observed in more detailed
models, fails to reproduce the observed rapid massive outflows and in
general has a much weaker effect on the host galaxy than predicted by
analytical calculations of feedback models.

On the other hand, a relatively simple improvement to the prescription
- injecting energy into a cone rather than a sphere - alleviates most
of these issues. The outflows become rapid, massive and close to
adiabatic, dense structures and inflowing filaments coexist with
outflowing gas, and gas fragmentation rates are significantly
increased. In particular, even extremely bright AGN are unable to shut
off gas accretion completely, allowing the SMBH to continue
growing. We suggest that this improvement could be used in
cosmological simulations in order to better model the effects that AGN
have on their host galaxies and clusters.

\section*{Acknowledgments}

KZ is funded by the Research Council of Lithuania grant
no. MIP-062/2013. MAB and SN acknowledge an STFC grant. MAB is funded
by a STFC research studentship.  We thank Justin Read for the use of
SPHS. This research used the DiRAC Complexity system, operated by the
University of Leicester IT Services, which forms part of the STFC
DiRAC HPC Facility (www.dirac.ac.uk).  This equipment is funded by BIS
National E-Infrastructure capital grant ST/K000373/1 and STFC DiRAC
Operations grant ST/K0003259/1. DiRAC is part of the UK National
E-Infrastructure.

\end{document}